\documentclass[10pt]{article}

\usepackage[english]{babel}
\usepackage{amsmath}
\usepackage{amssymb}
\usepackage{bm}
\usepackage{bbm}
\usepackage{mathrsfs,color}
\usepackage{graphics,graphicx,theorem}
\usepackage{dsfont}
\usepackage{setspace}

\usepackage{natbib}
\usepackage{multirow}
\usepackage{hhline}
\usepackage{hyperref}

\hypersetup{
	colorlinks=true,
	urlcolor=blue,
	citecolor=blue,
	linktoc=all,
	linkcolor=red} 

\usepackage[scriptsize]{subfigure}
\allowdisplaybreaks

\makeatletter
\renewcommand\@biblabel[1]{}
\makeatother

\def\bSig\mathbf{\Sigma}

\newcommand{\E}{\mathbb{E}}

\renewcommand{\P}{\mathbb{P}}

\newcommand{\qed}{$\square$}

\def\simiid{\stackrel{\mbox{\scriptsize{iid}}}{\sim}}

\usepackage[top=1.2in,bottom=1.2in,left=1.2in,right=1.2in]{geometry}

\begin{document}

\title{\bf {Rediscovery of Good-Turing estimators\\[-5pt] via Bayesian nonparametrics}}
\author{
Stefano Favaro\\
\normalsize{University of Torino and Collegio Carlo Alberto, Italy}\\
\normalsize{email: \texttt{stefano.favaro@unito.it}}
\bigskip\\
Bernardo Nipoti\\
\normalsize{University of Torino and Collegio Carlo Alberto, Italy}\\
\normalsize{email: \texttt{bernardo.nipoti@carloalberto.org}}
\bigskip\\
Yee Whye Teh\\
\normalsize{University of Oxford, UK}\\
\normalsize{email: \texttt{y.w.teh@stats.ox.ac.uk}}
}
\date{}
\maketitle
\thispagestyle{empty}

\setcounter{page}{1}
\begin{center}
\textbf{Abstract} 
\end{center}
The problem of estimating discovery probabilities originated in the context of statistical ecology, and in recent years it has become popular due to its frequent appearance in challenging applications arising in genetics, bioinformatics, linguistics, designs of experiments, machine learning, etc. A full range of statistical approaches, parametric and nonparametric as well as frequentist and Bayesian, has been proposed for estimating discovery probabilities. In this paper we investigate the relationships between the celebrated Good--Turing approach, which is a frequentist nonparametric approach developed in the 1940s, and a Bayesian nonparametric approach recently introduced in the literature. Specifically, under the assumption of a two parameter Poisson-Dirichlet prior, we show that Bayesian nonparametric estimators of discovery probabilities are asymptotically equivalent, for a large sample size, to suitably smoothed Good--Turing estimators. As a by-product of this result, we introduce and investigate a methodology for deriving exact and asymptotic credible intervals to be associated with the Bayesian nonparametric estimators of discovery probabilities. The proposed methodology is illustrated through a comprehensive simulation study and the analysis of Expressed Sequence Tags data generated by sequencing a benchmark complementary DNA library.

\vspace*{.2in}

\noindent\textsc{Keywords}: {Asymptotic equivalence; Bayesian nonparametrics; credible intervals; discovery probability; Expressed Sequence Tags; Good--Toulmin estimator; Good--Turing estimator; smoothing technique; two parameter Poisson-Dirichlet prior.} 

\maketitle


\section{Introduction} \label{s:intro}

Consider a population of individuals $(X_{i})_{i\geq1}$ belonging to an (ideally) infinite number of species $(X^{\ast}_{i})_{i\geq1}$ with unknown proportions $(p_{i})_{i\geq1}$. Given an initial observed sample of size $n$, a quantity of practical interest is the probability $D_{n,m}(l)$ of observing at the $(n+m+1)$-th drawn a species with frequency $l\geq0$ in the enlarged sample of size $n+m$, with the additional sample being unobserved. Formally, if $N_{i,n+m}$ denotes the frequency of $X^{\ast}_{i}$ in the enlarged sample, then
\begin{equation}\label{eq:main_disc}
D_{n,m}(l)=\sum_{i\geq1}p_{i}\mathds{1}_{\{l\}}(N_{i,n+m}).
\end{equation}
Clearly $D_{n,m}(0)$ corresponds to the proportion of yet unobserved species or, equivalently, the probability of discovering a new species at the $(n+m+1)$-th drawn. The random probability \eqref{eq:main_disc} is typically referred to as the $(m;l)$-discovery. While the $(0;l)$-discovery is of interest for estimating the probability of discovering new species or rare species, the $(m;l)$-discovery is typically of interest in decision problems regarding the size of the additional sample to collect.

A full range of statistical approaches, parametric and nonparametric as well as frequentist and Bayesian, have been proposed for estimating $D_{n,m}(l)$. These approaches have originally found applications in ecology, and their importance has grown considerably in recent years, driven by challenging applications arising in genetics, bioinformatics, linguistics, designs of experiments, machine learning, etc. See \citet{Bun(93)} and \cite{Bun(14)} for comprehensive reviews. In this paper we investigate the relationships between two approaches for estimating $D_{n,m}(l)$: i) the frequentist nonparametric approach which appeared in the seminal paper by \citet{Goo(53)}, and first developed by Alan M. Turing and Irving J. Good during their collaboration at Bletchley Park in the 1940s; ii) the Bayesian nonparametric approach recently introduced by \citet{Lij(07)} and \citet{Fav(12)}. In order to state our main contributions, we briefly review the relevant aspects of these two nonparametric approaches.

\subsection{The Good--Turing approach}

Let $\mathscr{H}$ be a parametric statistical hypothesis on the $p_{i}$'s, that is $\mathscr{H}$ determines the species composition of the population by specifying a distribution function over species and with a finite number of unknown parameters. Let $\boldsymbol{X}_{n}=(X_{1},\ldots,X_{n})$ be a random sample from $\mathscr{H}$, and let us denote by $M_{l,n}$ the number of species with frequency $l$ in $\boldsymbol{X}_{n}$. According to \citet{Goo(53)}, an estimator of $D_{n,0}(l)$ is $\check{\mathcal{D}}_{n,0}(l;\mathscr{H})=(l+1)\E_{\mathscr{H}}[M_{l+1,n+1}]/(n+1)$,  where $\E_{\mathscr{H}}$ denotes the expected value with respect to the distribution function specified by $\mathscr{H}$. For any $m\geq1$ let us consider the additional unobserved sample $(X_{n+1},\ldots,X_{n+m})$, and define $\gamma=m/n$. According to \citet{Goo(56)}, an estimator of $D_{n,m}(0)$ is $\check{\mathcal{D}}_{n,m}(0;\mathscr{H})=\sum_{i\geq1}(-\gamma)^{i-1}i\E_{\mathscr{H}}[M_{i,n+m}]/n$. Note that, in principle, $\E_{\mathscr{H}}[M_{l+1,n+1}]$ and $\E_{\mathscr{H}}[M_{i,n+m}]$ do not depend on the initial observed sample, unless the parameters characterizing $\mathscr{H}$ are estimated using such a  sample. Several examples of $\mathscr{H}$ are thoroughly discussed in \citet{Goo(53)} and, among them, we mention the Zipf-type distributions and the discretized Pearson distributions.

In order to dispense with the specification of the parametric statistical hypothesis $\mathscr{H}$, \citet{Goo(53)} proposed a large $n$ approximation of $\check{\mathcal{D}}_{n,0}(l;\mathscr{H})$ by replacing $\E_{\mathscr{H}}[M_{l+1,n+1}]/(n+1)$ with $m_{l+1,n}/n$, where $m_{l,n}$ denotes the number of species with frequency $l$ in the observed sample. In particular, if $x_{n}\bumpeq y_{n}$ means that $x_{n}$ is approximately equal to $y_{n}$ for large $n$, then we can write
\begin{equation}\label{eq:good_turing_1}
\check{\mathcal{D}}_{n,0}(l;\mathscr{H})\bumpeq\check{\mathcal{D}}_{n,0}(l)=(l+1)\frac{m_{l+1,n}}{n}.
\end{equation}
The large $n$ approximate estimator \eqref{eq:good_turing_1} is known as the Good--Turing estimator. A similar large $n$ approximation was proposed in \citet{Goo(56)} for $\check{\mathcal{D}}_{n,m}(0;\mathscr{H})$. Specifically, 
\begin{equation}\label{eq:good_toulmin_1}
\check{\mathcal{D}}_{n,m}(0;\mathscr{H})\bumpeq\check{\mathcal{D}}_{n,m}(0)=\frac{1}{n}\sum_{i\geq1}(-\gamma)^{i-1}im_{i,n}.
\end{equation}
$\check{\mathcal{D}}_{n,m}(0)$ is known as the Good--Toulmin estimator for the $(m;0)$-discovery. As observed by \citet{Goo(56)}, due to the alternating sign of the series which appears in the estimator \eqref{eq:good_toulmin_1}, if $\gamma$ is large then $\check{\mathcal{D}}_{n,m}(0)$ can yield inadmissible estimates. This instability arises even for values of $m$ moderately larger than $n$, typically $m$ greater than $n$ is enough for it to appear.

A peculiar feature of $\check{\mathcal{D}}_{n,0}(l)$ is that it depends on $m_{l+1,n}$, and not on $m_{l,n}$ as one would intuitively expect for an estimator of the $(0;l)$-discovery. Such a feature, combined with the irregular behaviour of the $m_{l,n}$'s for large $l$, makes $\check{\mathcal{D}}_{n,0}(l)$ a sensible approximation only if $l$ is sufficiently small with respect to $n$. Indeed for some large $l$ one might observe that $m_{l,n}>0$ and $m_{l+1,n}=0$, which provides the absurd estimate $\check{\mathcal{D}}_{n,0}(l)=0$, or that $m_{l,n}<m_{l+1,n}$ although the overall observed trend for $m_{l,n}$ is to decrease as $l$ increases. In order to overcome these drawbacks \citet{Goo(53)} suggested to smooth the irregular series of $m_{l,n}$'s into a more regular series to be used as a proxy. If $m_{l,n}^{\prime}$'s are the smoothed $m_{l,n}$'s with respect to a smoothing rule $\mathscr{S}$, then $\check{\mathcal{D}}_{n,0}(l;\mathscr{S})=(l+1)m_{l+1,n}^{\prime}/n$ is a more accurate approximation than $\check{\mathcal{D}}_{n,0}(l)$. Common  smoothing rules consider $m_{l,n}^{\prime}$, as a function of $l$, to be approximately parabolic or, alternatively, $m_{l,n}^{\prime}$ to be a certain proportion of the number of species in $\boldsymbol{X}_{n}$. An alternative method assumes $\mathscr{H}$ to be selected from a superpopulation with an assigned distribution. This flexible method was hinted at in \citet{Goo(53)} and then left as an open problem.

\subsection{The Bayesian nonparametric approach}

The approach in \citet{Lij(07)} and \citet{Fav(12)} is based on the randomization of  $p_{i}$'s. This is somehow reminiscent of the superpopulation smoothing hinted at by \cite{Goo(53)}. Specifically, let $P=\sum_{i\geq1}p_{i}\delta_{X^{\ast}_{i}}$ be a discrete random probability measure, namely $(p_{i})_{i\geq1}$ are nonnegative random weights such that $\sum_{i\geq1}p_{i}=1$ almost surely, and $(X^{\ast}_{i})_{i\geq1}$ are random locations independent of $(p_{i})_{i\geq1}$ and independent and identically distributed as a nonatomic distribution. The sample $\boldsymbol{X}_{n}$ is drawn from a population with species composition determined by $P$, i.e.
\begin{align}\label{eq:exchangeable_model}
X_i\,|\,P & \quad\simiid\quad P \qquad i=1,\ldots,n\\
\notag P & \quad\sim\quad \mathscr{P},
\end{align}
for any $n\geq1$, where $\mathscr{P}$ is a prior distribution over the species composition. Within the large class of priors considered in \citet{Lij(07)} and \citet{Fav(12)}, we focus on the two parameter Poisson-Dirichlet prior by \citet{Pit(95)}. Such a choice corresponds to set $p_1=V_1$ and $p_i=V_i\prod_{1\leq j\leq i-1}(1- V_j)$ where the $V_{j}$'s are independent Beta random variables with parameter $(1-\sigma,\theta+j\sigma)$, for any $\sigma\in(0,1)$ and $\theta>-\sigma$. We shorten ``two parameter Poisson-Dirichlet" by $\text{PD}(\sigma,\theta)$, and we denote by $P_{\sigma,\theta}$ a random probability measure distributed as $\text{PD}(\sigma,\theta)$ prior.

Under the framework \eqref{eq:exchangeable_model}, and with $\mathscr{P}$ being the $\text{PD}(\sigma,\theta)$ prior, \citet{Lij(07)} and \citet{Fav(12)} derived a Bayesian nonparametric estimator of the $(m;l)$-discovery. Specifically, let $\boldsymbol{X}_{n}$ be a sample from $P_{\sigma,\theta}$ featuring $K_{n}=k_{n}$ species with corresponding frequency counts $(M_{1,n},\ldots,M_{n,n})=(m_{1,n},\ldots,m_{n,n})$. From Proposition 2 in  \citet{Lij(07)}, the Bayesian nonparametric estimator of $D_{n,m}(0)$, with respect to a squared loss function, is
\begin{align}\label{eq:estim00}
\hat{\mathcal{D}}_{n,m}(0)=\frac{\theta+\sigma k_{n}}{\theta+n}\frac{(\theta+n+\sigma)_{m}}{(\theta+n+1)_{m}},
\end{align}
for any $m\geq0$, where $(a)_{n}=\prod_{0\leq i\leq n-1}(a+i)$ with the proviso $(a)_{0}\equiv 1$. For any $m\geq0$, let $(X_{n+1},\ldots,X_{n+m})$ be the additional unobserved sample from $P_{\sigma,\theta}$. According to Theorem 2 in \citet{Fav(12)}, the Bayesian nonparametric estimator of $D_{n,m}(l)$, with respect to a squared loss function, is
\begin{align}\label{eq:estim0l}
\hat{\mathcal{D}}_{n,m}(l)&=\sum_{i=1}^{l}{m\choose l-i}m_{i,n}(i-\sigma)_{l+1-i}\frac{(\theta+n-i+\sigma)_{m-l+i}}{(\theta+n)_{m+1}}\\
&\notag\quad+(1-\sigma)_{l}{m\choose l}(\theta+\sigma k_{n})\frac{(\theta+n+\sigma)_{m-l}}{(\theta+n)_{m+1}},
\end{align}
for any $l=1,\ldots,n+m$. According to the results displayed in \eqref{eq:estim00} and \eqref{eq:estim0l}, the Bayesian nonparametric approach has  two notable advantages with respect to the Good--Turing approach: i) it leads directly to exact estimators, thus avoiding the use of large $n$ approximations; ii) $\hat{\mathcal{D}}_{n,0}(l)$ is a function of $k_{n}$ and $m_{l,n}$, and not of $m_{l+1,n}$, thus avoiding the use of ad-hoc smoothing techniques to prevent absurd estimates determined by the irregular behavior of the $m_{l,n}$'s.

\subsection{Contributions of the paper and outline}

Let $a_{n}\simeq b_{n}$ mean that $\lim_{n\rightarrow+\infty}a_{n}/b_{n}=1$, namely $a_{n}$ and $b_{n}$ are asymptotically equivalent as $n$ tends to infinity. In this paper we show that the Bayesian nonparametric estimator $\hat{\mathcal{D}}_{n,0}(l)$ is asymptotically equivalent, as the sample size $n$ tends to infinity, to a Good--Turing estimator with suitably smoothed frequency counts. More precisely, for any $\sigma\in(0,1)$ we show that $\hat{\mathcal{D}}_{n,0}(l)\simeq\check{\mathcal{D}}_{n,0}(l;\mathscr{S}_{\text{PD}})$ as $n\rightarrow+\infty$, where $\mathscr{S}_{\text{PD}}$ is a smoothing rule such that $m_{l,n}$ is smoothed by
\begin{equation}\label{eq:smoother}
m_{l,n}^{\prime}=\frac{\sigma(1-\sigma)_{l-1}}{l!}k_{n}.
\end{equation}
While smoothing techniques were introduced  in \citet{Goo(53)} as an ad hoc tool for post processing the $m_{l}$'s in order to improve the performance of $\check{\mathcal{D}}_{n,0}(l)$, our result shows that, for a large sample size, a similar smoothing mechanism underlies the Bayesian framework \eqref{eq:exchangeable_model} with a $\text{PD}(\sigma,\theta)$ prior. We show that $\mathscr{S}_{\text{PD}}$ is related to the Poisson smoothing introduced in \citet{Goo(53)}, and we discuss a natural generalization of $\mathscr{S}_{\text{PD}}$ which leads to an interesting open problem.

Besides introducing an asymptotic relationship between $\hat{\mathcal{D}}_{n,0}(l)$ and $\check{\mathcal{D}}_{n,0}(l)$, we extend such a relationship to the $(m;l)$-discovery. Specifically, for any fixed $n$ and as $m$ tends to infinity, we show that $\hat{\mathcal{D}}_{n,m}(l)$ is asymptotically equivalent to a Good--Turing estimator $\check{\mathcal{D}}_{m,0}(l)$ in which $m_{l+1,m}$ is replaced by a smoothed version, via $\mathscr{S}_{\text{PD}}$, of the Bayesian nonparametric estimator $\hat{\mathcal{M}}_{n,m}(l+1)$ of the number of species with frequency $l$ in the enlarged sample. As a by-product of this result we introduce a methodology for deriving large $m$ asymptotic credible intervals for $\hat{\mathcal{D}}_{n,m}(l)$, thus completing the study in \citet{Lij(07)} and \citet{Fav(12)}. While the $\text{PD}(\sigma,\theta)$ prior leads to an explicit expression for the posterior distribution of $D_{n,m}(l)$, this expression involve combinatorial coefficients whose evaluation for large $m$ is cumbersome, thus preventing its implementation for determining exact credible intervals. Our methodology thus provides a fundamental tool in many situations of practical interest, arising especially in genomics, where $m$ is required to be very large and only a small portion of the population is sampled.

Our results are illustrated through a simulation study and the analysis of Expressed Sequence Tags (ESTs) data generated by sequencing a benchmark complementary DNA (cDNA) library. By means of a simulation study we compare $\check{\mathcal{D}}_{n,0}(l;\mathscr{S}_{\text{PD}})$ with smoothed Good--Turing estimators obtained by using the Poisson smoothing and a smoothing technique in \citet{Sam(01)}. Simulated data are generated from the Zeta distribution, whose power law behavior is common in numerous applications. In order to detect the effects of different smoothing techniques, we compare the smoothed Good--Turing estimators with $\check{\mathcal{D}}_{n,0}(l)$ and $\hat{\mathcal{D}}_{n,0}(l)$. A second numerical illustration is devoted to the large $m$ asymptotic credible intervals for the Bayesian nonparametric estimator $\hat{\mathcal{D}}_{n,m}(l)$. Using ESTs data we compare asymptotic confidence intervals for the Good--Toulmin estimator $\check{\mathcal{D}}_{n,m}(0)$ with asymptotic credible intervals for its  Bayesian nonparametric counterpart $\hat{\mathcal{D}}_{n,m}(0)$. This  study completes the numerical illustration presented in \citet{Fav(09)} and \citet{Fav(12)} on the same ESTs data.

In Section 2 we present and discuss the asymptotic equivalence between the Good--Turing approach and the Bayesian nonparametric approach under the assumption of the $\text{PD}(\sigma,\theta)$ prior. As a by-product of this asymptotic analysis, in Section 3 we introduce a methodology for associating large $m$ asymptotic credible intervals to $\hat{\mathcal{D}}_{n,m}(l)$. Section 4 contains numerical illustrations. Proofs of our results, as well as related additional materials, 
are postponed to the Appendix.


\section{Good--Turing estimators via Bayesian nonparametrics}\label{main_sec}

Under a $\text{PD}(\sigma,\theta)$ prior, the most notable difference between the Good--Turing estimator and its Bayesian nonparametric counterpart can be traced back to the different use of the information contained in the observed sample. As pointed out in the Introduction, $\check{\mathcal{D}}_{n,0}(0)$ is a function of $m_{1,n}$ while $\hat{\mathcal{D}}_{n,0}(0)$ in \eqref{eq:estim00} is a function of $k_{n}$. Furthermore, for any $l=1,\ldots,n$, $\check{\mathcal{D}}_{n,0}(l)$ is a function of $m_{l+1,n}$ while $\hat{\mathcal{D}}_{n,0}(l)$ in \eqref{eq:estim0l} is a function of $m_{l,n}$. In this section we show that, as $n$ tends to infinity, $\hat{\mathcal{D}}_{n,0}(l)$ is asymptotically equivalent to the smoothed Good--Turing estimator $\check{\mathcal{D}}_{n,0}(l;\mathscr{S}_{\text{PD}})$, where $\mathscr{S}_{\text{PD}}$ is the smoothing rule displayed in \eqref{eq:smoother}. A similar asymptotic equivalence, for fixed $n$ and as $m$ tends to infinity, holds between the estimators $\hat{\mathcal{D}}_{n,m}(l)$ and $\check{\mathcal{D}}_{m,0}(l)$. With a slight abuse of notation, throughout this section we write $X\,|\,Y$ to denote a random variable whose distribution coincides with the conditional distribution of $X$ given $Y$.

\subsection{Large $n$ asymptotic equivalences for $\hat{\mathcal{D}}_{n,0}(l)$}\label{sec:large}

We start by recalling the predictive distribution characterizing $P_{\sigma,\theta}$. Let $\boldsymbol{X}_{n}$ be a sample of size $n$ featuring $K_{n}=k_{n}$ species $X_{1}^{\ast},\ldots,X_{K_{n}}^{\ast}$ with frequencies $(N_{1,n},\ldots,N_{K_{n},n})=(n_{1,n},\ldots,n_{k_{n},n})$. According to the de Finetti's representation theorem, $\boldsymbol{X}_{n}$ is part of an exchangeable sequence $(X_{i})_{i\geq1}$ whose distribution has been characterized by \citet{Pit(95)} as follows
\begin{align}\label{eq:predictive}
\P[X_{n+1}\in\cdot\,|\,\boldsymbol{X}_{n}]=\frac{\theta+\sigma k_{n}}{\theta+n}\nu_{0}(\cdot)+\frac{1}{\theta+n}\sum_{i=1}^{k_{n}}(n_{i,n}-\sigma)\delta_{X_{i}^{\ast}}(\cdot), 
\end{align}
with $\nu_{0}$ being a nonatomic probability measure. The conditional probability \eqref{eq:predictive} is referred to as the predictive distribution of $P_{\sigma,\theta}$. Note that $\hat{\mathcal{D}}_{n,0}(l)$ can be read from \eqref{eq:predictive}, indeed from \eqref{eq:estim00} and \eqref{eq:estim0l} one has $\hat{\mathcal{D}}_{n,0}(0)=(\theta+\sigma k_{n})/(\theta+n)$ and $\hat{\mathcal{D}}_{n,0}(l)=(l-\sigma)m_{l,n}/(\theta+n)$, respectively. See \citet{Pit(95)} for details on \eqref{eq:predictive}, and on the joint distribution of $K_{n}$ and $(N_{1,n},\ldots,N_{K_{n},n})$ induced by \eqref{eq:predictive}.

The asymptotic equivalence between $\hat{\mathcal{D}}_{n,0}(l)$ and $\check{D}_{n,0}(l;\mathscr{S}_{\text{PD}})$ relies on an interesting interplay between the large $n$ asymptotic behaviors of $K_{n}$ and $M_{l,n}$ under a $\text{PD}(\sigma,\theta)$ prior. Specifically, let $A_{n}\stackrel{\text{a.s.}}{\simeq} B_{n}$ as $n\rightarrow+\infty$ mean that $\lim_{n\rightarrow+\infty}A_{n}/B_{n}=1$ almost surely, namely $A_{n}$ and $B_{n}$ are almost surely asymptotically equivalent as $n$ tends to infinity. By a direct application of Theorem 3.8 and Lemma 3.11 in \citet{Pit(06)}, one obtains the asymptotic equivalence
\begin{equation}\label{eq:prior_as_eq}
M_{l,n}\stackrel{\text{a.s.}}{\simeq}\frac{\sigma(1-\sigma)_{l-1}}{l!}K_{n}
\end{equation}
as $n\rightarrow+\infty$. In other terms, under a $\text{PD}(\sigma,\theta)$ prior, as the sample size $n$ tends to infinity the number of species with frequency $l$ becomes a proportion $\sigma(1-\sigma)_{l-1}/l!$ of the total number of species. We refer to the Appendix for additional details on \eqref{eq:prior_as_eq}. The next theorem combines \eqref{eq:predictive} and \eqref{eq:prior_as_eq} in order to establish the asymptotic equivalence between $\hat{\mathcal{D}}_{n,0}(l)$ and $\check{D}_{n,0}(l;\mathscr{S}_{\text{PD}})$.

\medskip

\textsc{Theorem 1.} Let $\boldsymbol{X}_{n}$ be a sample of size $n$ from $P_{\sigma,\theta}$ featuring $K_{n}=k_{n}$ species with corresponding frequency counts $(M_{1,n},\ldots,M_{n,n})=(m_{1,n},\ldots,m_{n,n})$. Then, as $n\rightarrow+\infty$, one has
\begin{equation}\label{eq:turing_bayes_pd}
\hat{\mathcal{D}}_{n,0}(l)\simeq(l+1)\frac{m_{l+1,n}}{n}\simeq(l+1)\frac{\frac{\sigma(1-\sigma)_{l}}{(l+1)!}k_{n}}{n}.
\end{equation}

\medskip

The smoothing rule $\mathscr{S}_{\text{PD}}$ clearly arises from the large $n$ asymptotic equivalence displayed in \eqref{eq:prior_as_eq}; indeed  $\mathscr{S}_{\text{PD}}$ smooths the frequency count $m_{l,n}$ by taking the proportion $\sigma(1-\sigma)_{l-1}/l!$ of $k_{n}$. Such a smoothing rule is somehow related to the Poisson smoothing $\mathscr{S}_{\text{Poi}}$, originally introduced by \citet{Goo(53)}, in which the frequency count $m_{l,n}$ is approximately equal to a proportion $\text{e}^{-\lambda}\lambda^{\tau+l-1}/(\tau+l-1)!$ of $k_{n}$, for any $\lambda>0$ and $\tau\geq0$ such that $\sum_{l\geq0}\check{D}_{n,0}(l;\mathscr{S}_{\text{Poi}})=1$. See Chapter 2 in \citet{Eng(78)} for a common example of Poisson smoothing where $\tau=1$ and $\lambda=n/k_{n}$. In particular $\mathscr{S}_{\text{PD}}$ is related to the Poisson smoothing corresponding to the choice $\tau=0$ and to a suitable randomization of the  parameter $\lambda$. Specifically, let us denote by $P_{\lambda}$ a discrete random variable with distribution $\P[P_{\lambda}=l]=\text{e}^{-\lambda}\lambda^{l-1}/(l-1)!$, that is the Poisson smoothing with $\tau=0$ and $\lambda>0$. If $G_{a,b}$ is Gamma random variable with parameter $(a,b)$ and $L_{\sigma}$ is a discrete random variable with distribution $\P[L_{\sigma}=l]=\sigma(1-\sigma)_{l-1}/l!$, then according to \citet{Dev(93)} $L_{\sigma}\stackrel{\text{d}}{=}1+P_{G_{1,1}G_{1,1-\sigma}/G_{1,\sigma}}$ where $G_{1,1}$, $G_{1,1-\sigma}$ and $G_{1,\sigma}$ are mutually independent.

A peculiar feature of the smoothing rule $\mathscr{S}_{\text{PD}}$ is that it depends only on $\sigma\in(0,1)$. This is because $\mathscr{S}_{\text{PD}}$ is obtained by suitably combining \eqref{eq:prior_as_eq}, which does not depend of the parameter $\theta$, with other two large $n$ asymptotic equivalences independent of $\theta$, namely: i) $\hat{\mathcal{D}}_{n,0}(0)\simeq \sigma k_{n}/n$ and ii) $\hat{\mathcal{D}}_{n,0}(l)\simeq (l-\sigma)m_{l,n}/n$. We conjecture that these asymptotic equivalences, as well as \eqref{eq:prior_as_eq}, hold for a more general class of priors considered in \citet{Lij(07)} and \citet{Fav(12)}. This is the class of Gibbs-type priors introduced by  \citet{Pit(03)} and including two of the most commonly used nonparametric priors, i.e., the $\text{PD}(\sigma,\theta)$ prior and the normalized generalized Gamma prior. See \citet{Deb(15)} for details. In other terms, our conjecture is that Theorem 1 holds for any Gibbs-type prior, that is the smoothing rule $\mathscr{S}_{\text{PD}}$ is invariant with respect to the choice of any prior in the Gibbs class. Intuitively, different smoothing rules for different Gibbs-type priors, if they exist, necessarily require to investigate the high-order large $n$ asymptotic behaviour of $\hat{\mathcal{D}}_{n,0}(l)$, and then combine it with a corresponding refinement of the asymptotic equivalence in \eqref{eq:prior_as_eq}. Work on this is ongoing.

\subsection{Large $m$ asymptotic equivalences for $\hat{\mathcal{D}}_{n,m}(l)$}

Let $\boldsymbol{X}_{n}$ be a sample of size $n$ from $P_{\sigma,\theta}$ featuring $K_{n}=k_{n}$ species with frequency counts $(M_{1,n},\ldots,M_{n,n})=(m_{1,n},\ldots,m_{n,n})$. For any $m\geq1$ let $(X_{n+1},\ldots,X_{n+m})$ be an additional unobserved sample. Let $K_{m}^{(n)}$ be the number of new species in $(X_{n+1},\ldots,X_{n+m})$ and let $M_{l,m}^{(n)}$ denote the number of species with frequency $l$ in $(X_{1},\ldots,X_{n+m})$. Since the additional sample is assumed to be not observed, let us introduce a randomized version of  $\hat{\mathcal{D}}_{n+m,0}(0)$ and $\hat{D}_{n+m,0}(l)$ as 
\begin{equation}\label{identity_0}
D_{0,m}^{(n)}=\frac{\theta+\sigma k_{n}+\sigma K_{m}^{(n)}}{\theta+n+m}
\end{equation}
and
\begin{equation}\label{identity_1}
D_{l,m}^{(n)}=(l-\sigma)\frac{M_{l,m}^{(n)}}{\theta+n+m},
\end{equation}
respectively. According to the expression \eqref{eq:estim00}, $K_{n}$ is a sufficient statistics for $\hat{\mathcal{D}}_{n,m}(0)$ and, therefore, the distribution of $D_{0,m}^{(n)}\,|\,\boldsymbol{X}_{n}$ takes on the interpretation of the posterior distribution, with respect to $\boldsymbol{X}_{n}$, of the $(m;0)$-discovery. Similarly, according to the expression \eqref{eq:estim0l},  $(K_{n},M_{1,n},\ldots,M_{l,n})$ is a sufficient statistic for $\hat{\mathcal{D}}_{n,m}(l)$ and, therefore, the distribution of $D_{n,m}^{(n)}(l)\,|\,\boldsymbol{X}_{n}$ takes on the interpretation of the posterior distribution, with respect to $\boldsymbol{X}_{n}$, of the $(m;l)$-discovery. 

By means of the identities introduced in \eqref{identity_0} and \eqref{identity_1}, the distribution of $D_{0,m}^{(n)}\,|\,\boldsymbol{X}_{n}$ and $D_{n,m}^{(n)}(l)\,|\,\boldsymbol{X}_{n}$ follows from the distribution of $K_{m}^{(n)}\,|\,\boldsymbol{X}_{n}$ and $M_{l,m}^{(n)}\,|\,\boldsymbol{X}_{n}$, respectively, which have been obtained in \citet{Lij(07)} and \citet{Fav(13)}. See the Appendix for details on these distributions. In particular, Proposition 1 in \citet{Fav(09)} showed that
\begin{displaymath}
\hat{\mathcal{K}}_{n,m}=\E[K_{m}^{(n)}\,|\,\boldsymbol{X}_{n}]=\frac{(\theta/\sigma+k_{n})}{(\theta+n)_{m}}\left((\theta+n+\sigma)_{m}-(\theta+n)_{m}\right),
\end{displaymath}
which is the Bayesian nonparametric estimator, with respect to a squared loss function, of $K_{m}^{(n)}$. Furthermore, for any $l=1,\ldots,n+m$, Proposition 7 in \citet{Fav(13)} showed that
\begin{align*}\label{estim_dist_freq}
\hat{\mathcal{M}}_{n,m}(l)=\E[M_{l,m}^{(n)}\,|\,\boldsymbol{X}_{n}]&=\sum_{i=1}^{l}{m\choose l-i}m_{i,n}(i-\sigma)_{l-i}\frac{(\theta+n-i+\sigma)_{m-l+i}}{(\theta+n)_{m}}\\
&\notag\quad+(1-\sigma)_{l-1}{m\choose l}(\theta+\sigma k_{n})\frac{(\theta+n+\sigma)_{m-l}}{(\theta+n)_{m}},
\end{align*}
which is the Bayesian nonparametric estimator, with respect to a squared loss function, of $M_{l,m}^{(n)}$. Note that, by means of \eqref{identity_0} and \eqref{identity_1} one obtains $\hat{\mathcal{D}}_{n,m}(0)=\E[D_{0,m}^{(n)}\,|\,\boldsymbol{X}_{n}]=(\theta+\sigma k_{n}+\sigma\hat{\mathcal{K}}_{n,m})/(\theta+n+m)$ and $\hat{\mathcal{D}}_{n,m}(l)=\E[D_{l,m}^{(n)}\,|\,\boldsymbol{X}_{n}]=(l-\sigma)\hat{\mathcal{M}}_{n,m}(l)/(\theta+n+m)$, which provides an alternative representation for the  estimators of the  $(m;0)$-discovery and $(m;l)$-discovery, respectively.

Similarly to Theorem 1, an asymptotic equivalence between $\hat{\mathcal{D}}_{n,m}(l)$ and $\check{\mathcal{D}}_{m,0}(l)$ relies on the interplay between the large $m$ asymptotic behaviors of the random variables $K_{m}^{(n)}\,|\,\boldsymbol{X}_{n}$ and $M_{l,m}^{(n)}\,|\,\boldsymbol{X}_{n}$. Specifically, for any $n\geq1$, by a direct application of Proposition 2 in \citet{Fav(09)} and Corollary 21 in \citet{Gne(07)} one obtains the following asymptotic equivalence
\begin{equation}\label{eq:prior_as_eq_post}
M^{(n)}_{l,m}\,|\,\boldsymbol{X}_{n}\stackrel{\text{a.s.}}{\simeq}\frac{\sigma(1-\sigma)_{l-1}}{l!}K_{m}^{(n)}\,|\,\boldsymbol{X}_{n}
\end{equation}
as $m\rightarrow+\infty$. In other terms, under a $\text{PD}(\sigma,\theta)$ prior, the large $m$ asymptotic equivalence between $M^{(n)}_{l,m}\,|\,\boldsymbol{X}_{n}$ and $K_{m}^{(n)}\,|\,\boldsymbol{X}_{n}$ coincides with the large $n$ asymptotic equivalence between $M_{l,n}$ and $K_{n}$. We refer to the Appendix for additional details on \eqref{eq:prior_as_eq_post}. The next theorem combines \eqref{identity_0}, \eqref{identity_1} and \eqref{eq:prior_as_eq_post} in order to establish an asymptotic equivalence between $\hat{\mathcal{D}}_{n,m}(l)$ and $\check{\mathcal{D}}_{m,0}(l)$.

\medskip

\textsc{Theorem 2.}  Let $\boldsymbol{X}_{n}$ be a sample of size $n$ from $P_{\sigma,\theta}$ featuring $K_{n}=k_{n}$  species with corresponding frequency counts $(M_{1,n},\ldots,M_{n,n})=(m_{1,n},\ldots,m_{n,n})$. Then, as $m\rightarrow+\infty$, one has
\begin{equation}\label{eq:toulmin_bayes_pd}
\hat{\mathcal{D}}_{n,m}(l)\simeq(l+1)\frac{\hat{\mathcal{M}}_{n,m}(l+1)}{m}\simeq(l+1)\frac{\frac{\sigma(1-\sigma)_{l}}{(l+1)!}\hat{\mathcal{K}}_{n,m}}{m}.
\end{equation}

\medskip

Besides discovery probabilities one is also interested in cumulative discovery probabilities, which are generalizations of the $(m;l)$-discovery defined as follows. For any $\tau\geq1$, let $\{l_{1},\ldots,l_{\tau}\}$ be a collection of distinct indexes such that $l_{i}\in\{0,1,\ldots,n+m\}$ for any $i=1,\ldots,\tau$. We define the $(m;l_{1},\ldots,l_{\tau})$-discovery as the cumulative discovery probability $D_{n,m}(l_{1},\ldots,l_{\tau})=\sum_{1\leq i\leq \tau}D_{n,m}(l_{i})$. Hence, the Bayesian nonparametric estimator of $(m;l_{1},\ldots,l_{\tau})$-discovery is
\begin{displaymath}
\hat{\mathcal{D}}_{n,m}(l_{1},\ldots,l_{\tau})=\sum_{i=1}^{\tau}\hat{\mathcal{D}}_{n,m}(l_{i}).
\end{displaymath}
Such a generalization of the $(m;l)$-discovery is mainly motivated by several applications of practical interest in which one aims at estimating the probability of discovering the so-called rare species. Specifically, these are species not yet observed or observed with a frequency smaller than a certain threshold $\tau$. Of course large $n$ and large $m$ asymptotic equivalences for the estimator $\hat{\mathcal{D}}_{n,m}(l_{1},\ldots,l_{\tau})$ follow by a direct application of Theorem 1 and Theorem 2, respectively.


\section{Credible intervals for $\hat{\mathcal{D}}_{n,m}(l_{1},\ldots,l_{\tau})$}\label{sec:cre}

While deriving the estimator $\hat{\mathcal{D}}_{n,m}(l)$, \citet{Lij(07)} and \citet{Fav(12)} did not consider the problem of associating a measure of uncertainty to $\hat{\mathcal{D}}_{n,m}(l)$. Such a problem reduces to the problem of evaluating the distribution of $D_{l,m}^{(n)}\,|\,\boldsymbol{X}_{n}$ by combining \eqref{identity_0} and \eqref{identity_1} with the distributions of $K_{m}^{(n)}\,|\,\boldsymbol{X}_{n}$ and $M_{l,m}^{(n)}\,|\,\boldsymbol{X}_{n}$ recalled in the Appendix. While the distribution of $D_{l,m}^{(n)}\,|\,\boldsymbol{X}_{n}$ is explicit, in many situations of practical interest the additional sample size $m$ is required to be very large and the computational burden for evaluating this posterior distribution becomes overwhelming. This happens, for instance, in various genomic applications where one has to deal with relevant portions of  cDNA libraries which typically consist of millions of genes. In this section we show how to exploit the large $m$ asymptotic behaviour of $D_{l,m}^{(n)}\,|\,\boldsymbol{X}_{n}$ in order to associate asymptotic credible intervals to the estimator $\hat{\mathcal{D}}_{n,m}(l)$.

Let $\boldsymbol{X}_{n}$ be a sample from $P_{\sigma,\theta}$ featuring $K_{n}=k_{n}$ species $X_{1}^{\ast},\ldots,X_{K_{n}}^{\ast}$ with frequencies summarized by the vector $(N_{1,n},\ldots,N_{K_{n},n})=(n_{1,n},\ldots,n_{k_{n},n})$. Let $Z_{\sigma,\theta,k_{n}}^{(n)}\stackrel{\text{d}}{=}B_{k_{n}+\theta/\sigma,n/\sigma-k_{n}}Z_{\sigma,(\theta+n)/\sigma}$ where $B_{a,b}$ is a Beta random variable with parameter $(a,b)$ and $Z_{\sigma,q}$ has density function $f_{Z_{\sigma,q}}(z)=\Gamma(q\sigma+1)z^{q-1-1/\sigma}f_{\sigma}(z^{-1/\sigma})/\sigma\Gamma(q+1)$, with $f_{\sigma}$ being the positive $\sigma$-stable density. By combining \eqref{identity_0} and \eqref{identity_1} with Proposition 2 in \citet{Fav(09)} and Corollary 21 in \citet{Gne(07)}, as $m\rightarrow+\infty$,
\begin{equation}\label{eq:asymp_dist_post}
\frac{D_{l,m}^{(n)}}{m^{\sigma-1}}\,|\,\boldsymbol{X}_{n}\stackrel{\text{a.s.}}{\longrightarrow}\frac{\sigma(1-\sigma)_{l}}{l!} Z_{\sigma,\theta,k_{n}}^{(n)}.
\end{equation}
For any $\tau\geq1$ and $\{l_{1},\ldots,l_{\tau}\}$ such that $l_{i}\in\{0,1,\ldots,n+m\}$ for any $i=1,\ldots,\tau$, let us introduce the random variable $D_{(l_{1},\ldots,l_{\tau}),m}^{(n)}=\sum_{1\leq i\leq\tau}D_{l_{i},m}^{(n)}$. The distribution of $D_{(l_{1},\ldots,l_{\tau}),m}^{(n)}\,|\,\boldsymbol{X}_{n}$ takes on the interpretation of the posterior distribution of the $(m;l_{1},\ldots,l_{\tau})$-discovery. In the next proposition we generalize the fluctuation limit \eqref{eq:asymp_dist_post} to the cumulative random probability $D_{(l_{1},\ldots,l_{\tau}),m}^{(n)}\,|\,\boldsymbol{X}_{n}$.

\medskip

\textsc{Proposition 1.} Let $\boldsymbol{X}_{n}$ be a sample of size $n$ from $P_{\sigma,\theta}$ featuring $K_{n}=k_{n}$  species with corresponding frequency counts $(M_{1,n},\ldots,M_{n,n})=(m_{1,n},\ldots,m_{n,n})$. Then, as $m\rightarrow+\infty$, one has
\begin{equation}\label{eq:limit_toulmin_cum_pd}
\frac{D_{(l_{1},\ldots,l_{\tau}),m}^{(n)}}{m^{\sigma-1}}\,|\,\boldsymbol{X}_{n}\stackrel{\text{w}}{\longrightarrow}\left(\sum_{i=1}^{\tau} \frac{\sigma(1-\sigma)_{l_{i}}}{l_{i}!}\right) Z_{\sigma,\theta,k_{n}}^{(n)}.
\end{equation}

\medskip

Fluctuation limits \eqref{eq:asymp_dist_post} and \eqref{eq:limit_toulmin_cum_pd} provide useful tools for approximating the distribution of $D_{l,m}^{(n)}\,|\,\boldsymbol{X}_{n}$ and $D_{(l_{1},\ldots,l_{\tau}),m}^{(n)}\,|\,\boldsymbol{X}_{n}$. The same fluctuation limits hold for any scaling factor $r(m)$ such that, as $m\rightarrow+\infty$, $r(m)\simeq m^{\sigma-1}$. This allows us to introduce a scaling factor finer than $m^{\sigma-1}$. Indeed it can be easily verified that, as soon as $\theta$ and $n$ are not overwhelmingly smaller than $m$, 
\begin{displaymath}
\hat{\mathcal{D}}_{n,m}^{\prime}(l)=m^{\sigma-1}\frac{\sigma(1-\sigma)_l}{l!}\E[Z^{(n)}_{\sigma,\theta,k_{n}}],
\end{displaymath}
with $\E[Z^{(n)}_{\sigma,\theta,k_{n}}]=(k_{n}+\theta/\sigma)\Gamma(\theta+n)/\Gamma(\theta+n+\sigma)$, can be far from $\hat{\mathcal{D}}_{n,m}(l)$. Hence, the corresponding asymptotic credible intervals could be far from the exact estimates. Of course the same issue appears for the estimator $\hat{\mathcal{D}}_{n,m}(l_{1},\ldots,l_{\tau})$. For this reason we consider the scaling factors $r^{\ast}(m,l)$ and $r^{\ast}(m,l_{1},\ldots,l_{\tau})$ in such a way that $\hat{\mathcal{D}}_{n,m}(l)=r^{\ast}(m,l)(\sigma(1-\sigma)_{l}/l!)\E[Z^{(n)}_{\sigma,\theta,k_{n}}]$ and  $\hat{\mathcal{D}}_{n,m}(l_1,\ldots,l_\tau)= r^{\ast}(m,l_{1},\ldots,l_{\tau})\sum_{1\leq i\leq \tau}(\sigma(1-\sigma)_{l_{i}}/l_{i}!)\E[Z^{(n)}_{\sigma,\theta,k_{n}}]$, and we define
\begin{equation}\label{eq:asymp_est_2}
\hat{\mathcal{D}}^{\ast}_{n,m}(l)=r^{\ast}(m,l)\frac{\sigma(1-\sigma)_l}{l!}\E[Z_{\sigma,\theta,n,k_{n}}]
\end{equation}
and
\begin{displaymath}
\hat{\mathcal{D}}^{\ast}_{n,m}(l_1,\ldots,l_\tau)=r^{\ast}(m,l_{1},\ldots,l_{\tau})\left(\sum_{i=1}^\tau\frac{\sigma(1-\sigma)_{l_i}}{l_i!}\right)\E[Z^{(n)}_{\sigma,\theta,k_{n}}].
\end{displaymath}
It can be easily verified that, as $m\rightarrow+\infty$, $r^{\ast}(m,l)\simeq m^{\sigma-1}$ and $r^{\ast}(m,l_{1},\ldots,l_{\tau})\simeq m^{\sigma-1}$. Explicit expressions of the scaling factors $r^{\ast}(m,l)$ and $r^{\ast}(m,l_{1},\ldots,l_{\tau})$ are provided in the Appendix. The reader is referred to  \citet{Fav(09)} for a similar approach in the context of Bayesian nonparametric inference for the number of new  species generated by the additional sample. 

We make use of \eqref{eq:asymp_dist_post} and \eqref{eq:limit_toulmin_cum_pd} for deriving large $m$ asymptotic credible intervals for $\hat{\mathcal{D}}_{n,m}(l)$ and $\hat{\mathcal{D}}_{n,m}(l_1,\ldots,l_\tau)$. This can be readily done by evaluating appropriate quantiles of the distribution of  $Z^{(n)}_{\sigma,\theta,k_{n}}$. For instance let $s_{1}$ and $s_{2}$ be quantiles of the distribution of $Z^{(n)}_{\sigma,\theta,k_{n}}$ such that $(s_1,s_2)$ is the 95\% credible interval with respect to this distribution. Then, $(r^{\ast}(m,l)\sigma(1-\sigma)_{l}s_1/l!,r^{\ast}(m,l)\sigma(1-\sigma)_{l}s_2/l!)$ is a 95\% asymptotic credible interval for $\hat{\mathcal{D}}_{n,m}(l)$. Analogous observations hold true for the estimator $\hat{\mathcal{D}}_{n,m}(l_1,\ldots,l_\tau)$. In order to determine the quantiles $s_{1}$ and $s_{2}$, we resort to a simulation algorithm for sampling the limiting random variable $Z^{(n)}_{\sigma,\theta,k_{n}}$. Note that, according to the definition of $Z^{(n)}_{\sigma,\theta,k_{n}}$, this procedure involves sampling from the random variable $Z_{\sigma,q}$ with density function $f_{Z_{\sigma,q}}(z)=\Gamma(q\sigma+1)z^{q-1-1/\sigma}f_{\sigma}(z^{-1/\sigma})/\sigma\Gamma(q+1)$.

A strategy for sampling $Z_{\sigma,q}$ was proposed by \citet{Fav(09)}. Specifically, let $L_{\sigma,q}=Z_{\sigma,q}^{-1/\sigma}$ and we introduce a Gamma random variable $U_q$ with parameter $(q,1)$. Then, conditionally on $U_q=u$, the distribution of $L_{\sigma,q}$ has density function proportional to $f_\sigma(x)\exp\{-ux\}$. Therefore, the problem of sampling from $Z_{\sigma,q}$ boils down to the problem of sampling from an exponentially tilted stable distribution. Here we improve the sampling scheme proposed in \citet{Fav(09)} by resorting to the fast rejection algorithm recently proposed in \citet{Hof(11)} for sampling from an exponentially tilted positive $\sigma$-stable random variable. Summarizing, in order to generate random variates from the distribution of $Z^{(n)}_{\sigma,\theta,k_{n}}$, we have the following steps: i) sample $B_{k_{n}+\theta/\sigma,n/\sigma-k_{n}}$; ii) sample $G_{(\theta+n)/\sigma,1}$ and set $U_{(\theta+n)/\sigma}=G_{(\theta+n)/\sigma,1}^{1/\sigma}$; iii) given $U_{(\theta+n)/\sigma}=u$, sample $L_{\sigma,(\theta+n)/\sigma}$ from density proportional to $f_\sigma(x)\exp\{-ux\}$, by means of the fast rejection sampling, and set $Z_{\sigma,(\theta+n)/\sigma}=L_{\sigma,(\theta+n)/\sigma}^{-\sigma}$; iv) set $Z^{(n)}_{\sigma,\theta,k_{n}}=B_{k_{n}+\theta/\sigma,n/\sigma-k_{n}}Z_{\sigma,(\theta+n)/\sigma}$.


\section{Illustrations}\label{sec:ill}

In order to implement our results, the first issue to be faced is the specification of the parameter $(\sigma,\theta)$ in the $\text{PD}(\sigma,\theta)$ prior. Hereafter, following the approach of \citet{Lij(07)} and \citet{Fav(12)}, we resort to an empirical Bayes procedure. Specifically let $\boldsymbol{X}_{n}$ be a sample from $P_{\sigma,\theta}$ featuring $K_{n}=k_{n}$  species with frequencies $(N_{1,n},\ldots,N_{K_{n},n})=(n_{1,n},\ldots,n_{k_{n},n})$. The empirical Bayes procedure consists in choosing $\theta$ and $\sigma$ that maximize the distribution of $\boldsymbol{X}_{n}$. This, under a $\text{PD}(\sigma,\theta)$ prior, corresponds to setting $(\sigma,\theta)=(\hat\sigma,\hat\theta)$, where
\begin{equation}\label{empirical}
(\hat\sigma,\hat\theta)=\operatorname*{arg\,max}_{(\sigma,\theta)}\left\{\frac{\prod_{i=0}^{k_{n}-1}(\theta+i\sigma)}{(\theta)_{n}}\prod_{i=1}^{k_{n}}(1-\sigma)_{n_{i,n}-1}\right\}.
\end{equation}
One could also specify a prior distribution on the parameter $(\sigma,\theta)$ and then seek a full Bayesian inference.  However, in terms of estimating $D_{n,m}(l)$, there are no relevant differences between this fully Bayes approach and the empirical Bayes approach, given the posterior distribution of $(\sigma,\theta)$ is highly concentrated; this is typically the case of large datasets since the parameter $(\sigma,\theta)$ directly describe the distribution of the observables. See Section \ref{sec:CI} for a more detailed discussion on these aspects. In the sequel, in order to keep the exposition as simple as possible, we consider the specification of $(\sigma,\theta)$ via the empirical Bayes procedure \eqref{empirical}.

\subsection{A comparative study for $\hat{\mathcal{D}}_{n,0}(l)$, $\check{\mathcal{D}}_{n,0}(l)$ and $\check{\mathcal{D}}_{n,0}(l;\mathscr{S})$}\label{sec:comp}
We compare the performance of the Bayesian nonparametric estimators for the $(0;l)$-discovery with respect to the corresponding Good--Turing estimators and smoothed Good--Turing estimators, for some choices of the smoothing rule. We draw 500 samples of size $n=1000$ from a Zeta distribution with scale parameter $s=1.5$. Recall that a Zeta random variable $Z$ is such that $\P[Z=z]=z^{-s}/C(s)$ where $C(s)=\sum_{i\geq 1}i^{-s}$, for $s>1$. Next we order the samples according to the number of observed distinct species ${k_n}$ and we split them in 5 groups. Specifically, for $i=1,2,\ldots,5$, the $i$-th group of samples will be composed by 100 samples featuring a total number of observed distinct species ${k_n}$ that stays between the quantiles of order $(i-1)/5$ and $i/5$ of the empirical distribution of $k_{n}$. We therefore pick at random one sample for each group and label it with the corresponding index $i$. This procedure leads to a total number of 5 samples of 1000 observations with different species compositions.

We use these simulated datasets for comparing estimators for the $(0;l)$-discovery with the true value of $D_{n,0}(l)$, for $l=0,1,5,10,20,30$. Specifically, we consider the Bayesian nonparametric estimator $\hat{\mathcal{D}}_{n,0}(l)$, the Good--Turing estimator $\check{\mathcal{D}}_{n,0}(l)$, the smoothed Good--Turing estimator $\check{\mathcal{D}}_{n,0}(l;\mathscr{S}_{\text{PD}})$, and the Poisson smoothed Good--Turing estimator $\check{\mathcal{D}}_{n,0}(l;\mathscr{S}_\text{Poi})$ with $\tau=1$ and $\lambda=n/k_{n}$. Finally, we also consider the so-called Simple Good--Turing estimator, denoted by $\check{\mathcal{D}}_{n,0}(l;\mathscr{S}_\text{SGT})$, which is a popular smoothed Good--Turing estimator discussed in Chapter 7 of \citet{Sam(01)}. Specifically, in the Simple Good--Turing estimator the smoothing rule $\mathscr{S}_\text{SGT}$ consists in first computing, for large $l$, some values $z_{l,n}$ that take into account both the positive frequency counts $m_{l,n}$ and the surrounding zero values, and then in resorting to a line of best fit for the pairs $\left(\log_{10}(l),\log_{10}(z_{l,n})\right)$ in order to obtain the smoothed values $m_{l,n}^\prime$.

\begin{table}[h!p!t!]
      \caption{Simulated data from a Zeta distribution. Comparison between the true $(0;l)$-discovery $D_{n,0}(l)$ with the estimate obtained by $\hat{\mathcal{D}}_{n,0}(l)$, $\check{\mathcal{D}}_{n,0}(l)$, $\check{\mathcal{D}}_{n,0}(l;\mathscr{S}_{\text{Poi}})$, $\check{\mathcal{D}}_{n,0}(l;\mathscr{S}_{\text{PD}})$ and $\check{\mathcal{D}}_{n,0}(l;\mathscr{S}_{\text{SGT}})$.}
      \label{table_samples}
  \centering{
     \small{
      \begin{tabular}{c|c|ccccc}
      \hline
   & Sample & 1 & 2 & 3 & 4 & 5\\ 
      \hline
      &$k_n$ & 136 & 139 & 141 & 146 & 155\\
      &$\hat\sigma$ & 0.6319 & 0.6710 & 0.7107 & 0.6926 & 0.6885\\
      &$\hat\theta$  & 1.2716 & 0.6815 & 0.2334 & 0.5000 & 0.7025\\
      \hline
      \multirow{6}{*}{$l=0$} &$D_{n,0}(l)$ & 0.0984 & 0.0997 & 0.0931 & 0.0924 & 0.0927\\
      & $\hat{\mathcal{D}}_{n,0}(l)$ & 0.0871 & 0.0939 & 0.1004 & 0.1016 & 0.1073\\
      & $\check{\mathcal{D}}_{n,0}(l)$ & 0.0870 & 0.0950 & 0.1040 & 0.1040 &  0.1080\\
      & $\check{\mathcal{D}}_{n,0}(l;\mathscr{S}_{\text{Poi}})$  & 0.0006 & 0.0008 & 0.0008 & 0.0011 & 0.0016\\
       & $\check{\mathcal{D}}_{n,0}(l;\mathscr{S}_{\text{PD}})$ & 0.0859 & 0.0933 & 0.1002 & 0.1011 &  0.1067\\
      & $\check{\mathcal{D}}_{n,0}(l;\mathscr{S}_{\text{SGT}})$  & 0.0870 & 0.0950 & 0.1040 & 0.1040 &  0.1080\\[6pt]
      \multirow{6}{*}{$l=1$}&$D_{n,0}(l)$  & 0.0273 & 0.0272 & 0.0478 & 0.0365 &  0.0331\\
      & $\hat {\mathcal{D}}_{n,0}(l)$ & 0.0320 & 0.0312 & 0.0301 & 0.0319 & 0.0336\\
      &$\check {\mathcal{D}}_{n,0}(l)$ & 0.0320 & 0.0220 & 0.0160 & 0.0240 & 0.0300\\
      & $\check {\mathcal{D}}_{n,0}(l;\mathscr{S}_{\text{Poi}})$ & 0.0047 & 0.0054 & 0.0059 & 0.0073 &  0.0102\\
        & $\check {\mathcal{D}}_{n,0}(l;\mathscr{S}_{\text{PD}})$ & 0.0316 & 0.0307 & 0.0290 & 0.0311 & 0.0332\\
      & $\check {\mathcal{D}}_{n,0}(l;\mathscr{S}_{\text{SGT}})$ & 0.0319 & 0.0221 & 0.0161 & 0.0240 & 0.0300\\[6pt]
      \multirow{6}{*}{$l=5$}&$D_{n,0}(l)$ & 0.0060 & 0.0238 & 0.0132 & 0.0154 & 0.0046\\
      & $\hat {\mathcal{D}}_{n,0}(l)$ & 0.0044 & 0.0173 & 0.0086 & 0.0215 & 0.0043\\
      &$\check {\mathcal{D}}_{n,0}(l)$& 0.0240 & 0.0180 & 0.0120 & 0.0180 & 0.0120\\
      & $\check {\mathcal{D}}_{n,0}(l;\mathscr{S}_{\text{Poi}})$ & 0.1148 & 0.1206 & 0.1243 & 0.1332 & 0.1470\\
       & $\check {\mathcal{D}}_{n,0}(l;\mathscr{S}_{\text{PD}})$ & 0.0126 & 0.0114 & 0.0101 & 0.0111 & 0.0120\\
      & $\check {\mathcal{D}}_{n,0}(l;\mathscr{S}_{\text{SGT}})$ & 0.0044 & 0.0176 & 0.0089 & 0.0219 & 0.0044\\[6pt]
      \multirow{6}{*}{$l=10$}&$D_{n,0}(l)$ & 0.0105 & 0 & 0.0105 & 0.0092 & 0.0202\\
      & $\hat {\mathcal{D}}_{n,0}(l)$  & 0.0094 & 0 & 0.0093 & 0.0093 & 0.0186\\
      &$\check {\mathcal{D}}_{n,0}(l)$& 0 & 0 & 0.0220 & 0.0110 & 0.0110\\
      & $\check {\mathcal{D}}_{n,0}(l;\mathscr{S}_{\text{Poi}})$ & 0.0816 & 0.0769 & 0.0738 & 0.0664 & 0.0543\\
      & $\check {\mathcal{D}}_{n,0}(l;\mathscr{S}_{\text{PD}})$ & 0.0082 & 0.0072 & 0.0062 & 0.0070 & 0.0075\\
      & $\check {\mathcal{D}}_{n,0}(l;\mathscr{S}_{\text{SGT}})$ & 0.0093 & 0 & 0.0094 & 0.0093 & 0.0186\\[6pt]
      \multirow{6}{*}{$l=20$}&$D_{n,0}(l)$ & 0 & 0.0142 & 0.0169 & 0 & 0\\
      & $\hat {\mathcal{D}}_{n,0}(l)$ & 0 & 0.0193 & 0.0193 & 0 & 0\\
      &$\check {\mathcal{D}}_{n,0}(l)$& 0 & 0 & 0 & 0 & 0\\
      & $\check {\mathcal{D}}_{n,0}(l;\mathscr{S}_{\text{Poi}})$ & 0.0001 & 0.0000 & 0.0000 & 0.0000 & 0.0000\\
      & $\check {\mathcal{D}}_{n,0}(l;\mathscr{S}_{\text{PD}})$ & 0.0053 & 0.0046 & 0.0038 & 0.0043 & 0.0047\\
      & $\check {\mathcal{D}}_{n,0}(l;\mathscr{S}_{\text{SGT}})$ & 0 & 0.0194 & 0.0195 & 0 & 0\\[6pt]
      \multirow{6}{*}{$l=30$}&$D_{n,0}(l)$ & 0.0260 & 0 & 0 & 0 & 0\\
      & $\hat {\mathcal{D}}_{n,0}(l)$ & 0.0293 & 0 & 0 & 0 & 0\\
      &$\check {\mathcal{D}}_{n,0}(l)$& 0 & 0 & 0 & 0 & 0.0310\\
      & $\check {\mathcal{D}}_{n,0}(l;\mathscr{S}_{\text{Poi}})$ & 0.0000 & 0.0000 & 0.0000 & 0.0000 & 0.0000\\
      & $\check {\mathcal{D}}_{n,0}(l;\mathscr{S}_{\text{PD}})$ & 0.0041 & 0.0035 & 0.0029 & 0.0033 & 0.0036\\
      & $\check {\mathcal{D}}_{n,0}(l;\mathscr{S}_{\text{SGT}})$ & 0.0292 & 0 & 0 & 0 & 0\\[3pt]
\hline
       \multicolumn{2}{c}{$\text{SSE}(\hat{\mathcal{D}}_{n,0})$}& 0.0006 & 0.0016 & 0.0007 & 0.0007 & 0.0006\\
       \multicolumn{2}{c}{$\text{SSE}(\check{\mathcal{D}}_{n,0})$}& 0.3475 & 0.3773 & 0.3460 & 0.3575 & 0.3530\\
       \multicolumn{2}{c}{$\text{SSE}(\check{\mathcal{D}}_{n,0}(\mathscr{S}_{\text{Poi}}))$}& 0.2657 & 0.2723 & 0.2765 & 0.2769 & 0.2745\\
       \multicolumn{2}{c}{$\text{SSE}(\check{\mathcal{D}}_{n,0}(\mathscr{S}_{\text{PD}}))$}& 0.1748 & 0.1748 & 0.1753 & 0.1746 & 0.1747\\
       \multicolumn{2}{c}{$\text{SSE}(\check{\mathcal{D}}_{n,0}(\mathscr{S}_{\text{SGT}}))$} & 0.0007 & 0.0018 & 0.0014 & 0.0008 & 0.0007\\[3pt]
       \hline
      \end{tabular}}}
\end{table}

Table \ref{table_samples} summarizes the result of our comparative study. As an overall measure for the performance of the estimators, we use the sum of squared error (SSE) defined, for a generic estimator $\hat D(l)$ of the $(0,l)$-discovery, as $\text{SSE}(\hat D(l))=\sum_{0\leq l\leq n}(\hat D(l)-d_{n,0}(l))^2$, with $d_{n,0}(l)$ being the true value of $D_{n,0}(l)$. By looking at the SSE in Table \ref{table_samples} it is apparent that $\hat{\mathcal{D}}_{n,0}(l)$ and $\check{\mathcal{D}}_{n,0}(l;\mathscr{S}_\text{SGT})$ are much more accurate than the others. As expected, the Good--Turing estimator $\check{\mathcal{D}}_{n,0}(l)$ has a good performance only for small values of $l$, while inconsistencies arise for large frequencies thus explaining the amplitude of the resulting SSE. For instance, since sample $i=3$ features one species that has frequency $l=20$ and no species with frequency $l=21$, the Good--Turing estimator $\check{\mathcal{D}}_{n,0}(20)$ gives $0$ while, clearly, there is positive probability to observe the species appeared $20$ times in the sample. Finally, $\check{\mathcal{D}}_{n,0}(l;\mathscr{S}_{\text{PD}})$ yields a smaller SSE than $\check{\mathcal{D}}_{n,0}(l;\mathscr{S}_\text{Poi})$. However, the poor accuracy of $\check{\mathcal{D}}_{n,0}(l;\mathscr{S}_{\text{PD}})$ and $\check{\mathcal{D}}_{n,0}(l;\mathscr{S}_\text{Poi})$, compared to $\hat{\mathcal{D}}_{n,0}(l)$  and $\check{\mathcal{D}}_{n,0}(l;\mathscr{S}_\text{SGT})$, shows that the parametric assumptions underlying the smoothing rules $\mathscr{S}_{\text{Poi}}$ and $\mathscr{S}_{\text{PD}}$ are not suitable for data generated according to a Zeta distribution.

\subsection{Credible intervals for $\hat{\mathcal{D}}_{n,m}(l_{1},\ldots,l_{\tau})$}\label{sec:CI}

We illustrate the implementation of the asymptotic credible intervals for the Bayesian nonparametric estimator $\hat{\mathcal{D}}_{n,m}(l_{1},\ldots,l_{\tau})$ through the analysis of ESTs data generated by sequencing a benchmark cDNA library. ESTs represent an efficient way to characterize expressed genes from an organism. The rate of gene discovery depends on the degree of redundancy of the cDNA library from which such sequences are obtained. Correctly estimating the relative redundancy of such libraries, as well as other quantities such as the probability of sampling a new or a rarely observed gene, is of fundamental importance since it allows one to optimize the use of expensive experimental sampling techniques.  Hereafter we consider the \emph{Naegleria gruberi} cDNA libraries prepared from cells grown under different culture conditions, namely aerobic and anaerobic. See \citet{Sus(04)} for additional details.

\begin{figure}[h!p!t!]
\centering
\subfigure[Naegleria Aerobic]{
\includegraphics[width=0.48\linewidth]{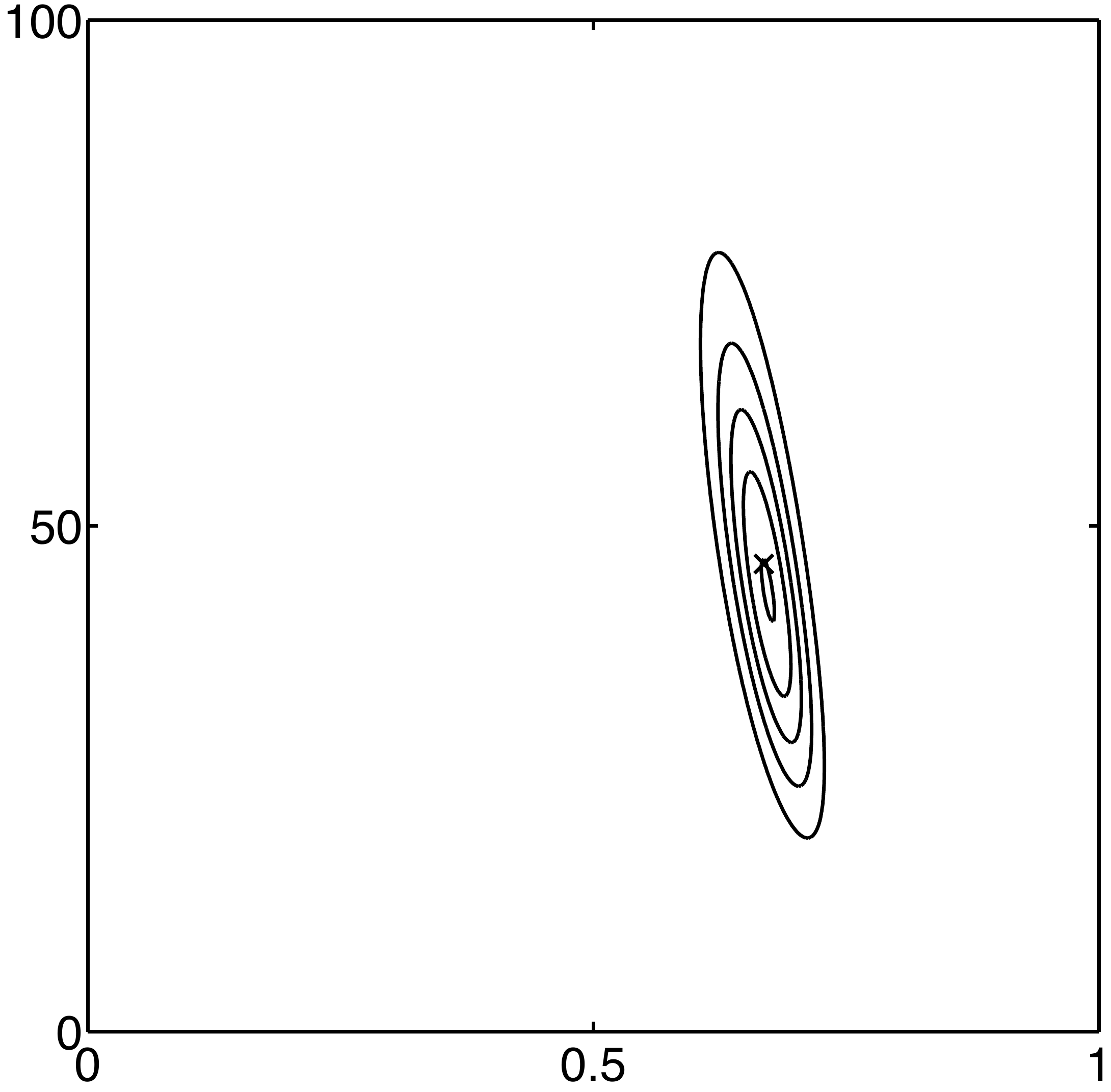}}
\subfigure[Naegleria Anaerobic]{
\includegraphics[width=0.48\linewidth]{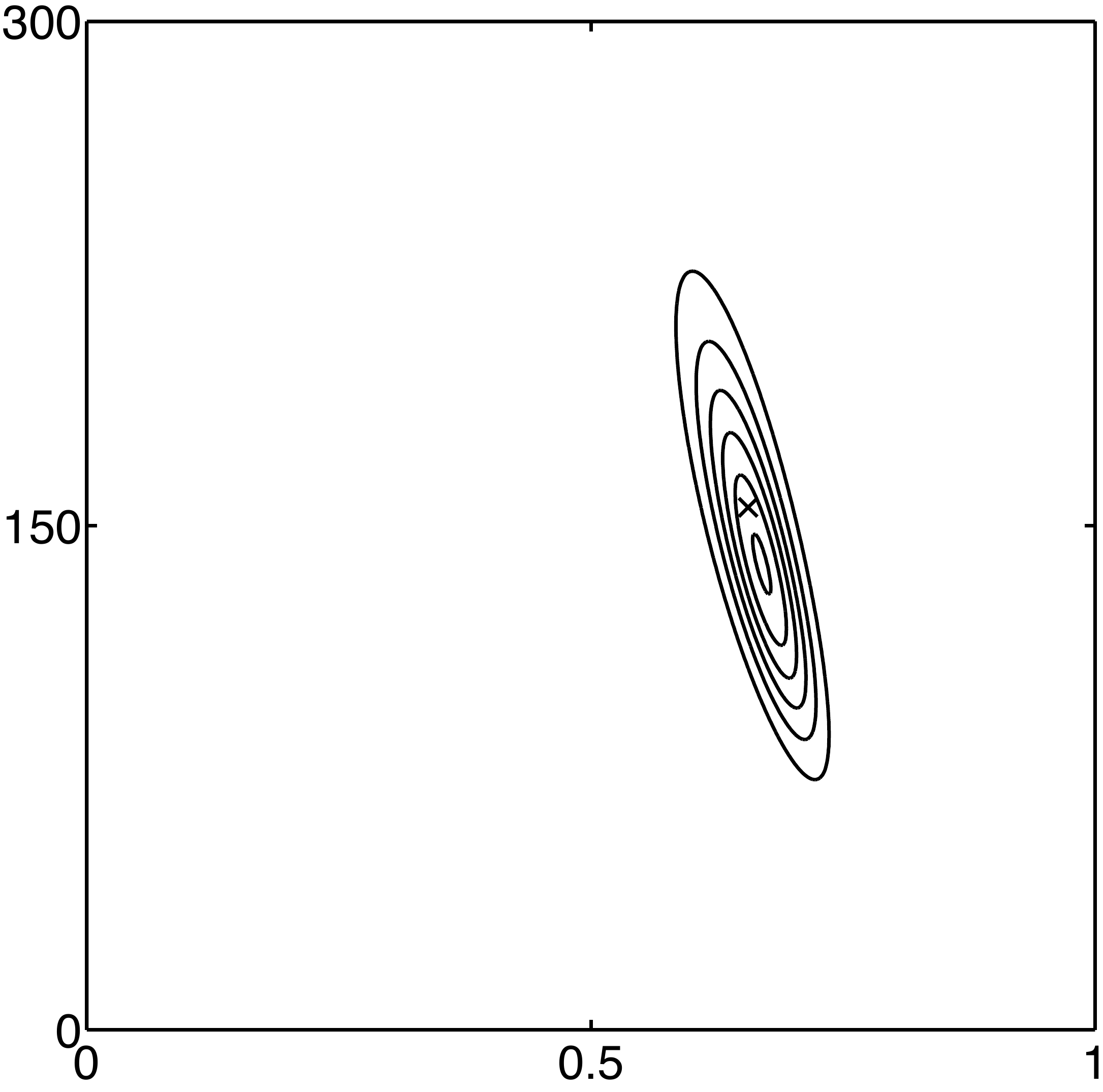}}
\caption{Contour lines of the posterior distribution of the parameter $(\sigma,\theta)$. The cross marks denote the estimates $(\hat\sigma,\hat\theta)$ obtained by means of the empirical Bayes procedure \eqref{empirical}.}
\label{fig:FB_vs_EB}
\end{figure}

The \emph{Naegleria gruberi} aerobic library consists of $n=959$ ESTs with $k_{n}=473$ distinct genes and $m_{i,959}=346, 57, 19, 12, 9, 5, 4, 2, 4, 5, 4, 1, 1, 1, 1, 1, 1$, for $i=\{1,2,\ldots,12\}\cup\{16,17,18\}\cup\{27\}\cup\{55\}$. The \emph{Naegleria gruberi} anaerobic library consists of $n=969$ ESTs with $k_{n}=631$ distinct genes and $m_{i,969}=491, 72, 30, 9, 13, 5, 3, 1, 2, 0, 1, 0, 1$, for $i\in\{1,2,\ldots,13\}$. A fully Bayesian approach involves the specification of a prior distribution for the parameter $(\sigma,\theta)$. Let us consider independent priors for $\sigma$ and $\theta$, namely a Uniform distribution on $(0,1)$ for $\sigma$ and a Gamma distribution with shape parameter $1$ and scale parameter $100$, for $\theta$. Figure \ref{fig:FB_vs_EB} shows  the contour lines of the posterior distribution of $(\sigma,\theta)$; note that these posterior distributions are rather concentrated on a small range of values for $\sigma$. The empirical Bayes approach \eqref{empirical} lead to the following estimates for $(\sigma,\theta)$: $(\hat{\sigma},\hat{\theta})=(0.669, 46.241)$ for the \emph{Naegleria gruberi} aerobic library and $(\hat{\sigma},\hat{\theta})=(0.656, 155.408)$ for the \emph{Naegleria gruberi} anaerobic library. These values are very close to the mode of the corresponding posterior distributions. See the cross marks in Figure \ref{fig:FB_vs_EB}. As a matter of fact, the fully Bayesian approach and the empirical Bayes approach lead to very similar estimates for $D_{n,m}(l)$. For instance, by adopting both the empirical Bayes approach and the fully Bayesian approach we get $\hat{\mathcal{D}}_{n,0}(0)=0.36$ for the \emph{Naegleria gruberi} aerobic library and $\hat{\mathcal{D}}_{n,0}(0)=0.51$ for the \emph{Naegleria gruberi} anaerobic library. This observation supports our choice of undertaking the empirical Bayes approach \eqref{empirical}. The reader is referred to the Appendix for a sensitivity analysis of the asymptotic credible intervals for $\hat{\mathcal{D}}_{n,m}(0)$, with respect to the choice of the parameter $(\sigma,\theta)$.

We now focus on the \emph{Naegleria gruberi} aerobic library, and observe that the estimates of the $(m;l)$-discovery provided by the exact estimator $\hat{\mathcal{D}}_{n,m}(0)$, for $m=n,10n,100n$, are $0.289,0.165,0.080$, respectively, while the corresponding estimates provided by the asymptotic estimator  $\hat{\mathcal{D}}_{n,m}^{\prime}(0)$ gives $0.367,0.171,0.080$. It is apparent that $\hat{\mathcal{D}}_{n,m}^{\prime}(0)$ provides estimates that are close to the exact estimates only when $m$ is very large. This motivates the use of asymptotic estimator $\hat{\mathcal{D}}^{\ast}_{n,m}(0)$ with a more accurate scaling factor. Similar considerations hold for the \emph{Naegleria gruberi} anaerobic library. This comparative study between the asymptotic estimators $\hat{\mathcal{D}}_{n,m}^{\prime}(0)$ and $\hat{\mathcal{D}}^{\ast}_{n,m}(0)$, as well as the corresponding credible intervals, is presented in Table \ref{table_scaling}.

\begin{table}[h!p!t!]
\caption{\label{table_scaling}\emph{Naegleria} aerobic and \emph{Naegleria} anaerobic libraries. Comparison between $\hat{\mathcal{D}}_{n,m}(0)$ and the corresponding asymptotic estimators $\hat{\mathcal{D}}_{n,m}^{\prime}(0)$ and $\hat{\mathcal{D}}^{\ast}_{n,m}(0)$. For the asymptotic estimators $95\%$ credible intervals (c.i.) are provided.\medskip}
\centering{
\tabcolsep=0.2cm
\footnotesize{
\begin{tabular}{*{7}{c}}
\hline
Library & $m$ & $\hat{\mathcal{D}}_{n,m}(0)$  &  \multicolumn{2}{c}{rate $m^{\sigma-1}$} & \multicolumn{2}{c}{rate $r^{\ast}(m,0)$}\\[0.1cm]
\cline{4-5}\cline{6-7}\\[-0.2cm]
& &  & $\hat{\mathcal{D}}_{n,m}^{\prime}(0)$ & $95\%$ c.i. &$\hat{\mathcal{D}}^{\ast}_{n,m}(0)$ & $95\%$ c.i.\\[0.1cm]\hline\\[-0.2cm]
\emph{Naegleria} Aerobic&$n$&0.289&0.367&$(0.339, 0.395)$&0.289&$(0.267, 0.312)$\\[0.2cm]
$(n=959)$&10$n$&0.165&0.171&$(0.158, 0.184)$&0.165&$(0.153, 0.178)$\\[0.2cm]
 &100$n$&0.080&0.080&$(0.074, 0.086)$&0.080&$(0.073, 0.086)$\\[0.2cm]
\emph{Naegleria} Anaerobic&$n$&0.409&0.533&$(0.505, 0.561)$&0.409&$(0.387, 0.431)$\\[0.2cm]
$(n=969)$&10$n$&0.232&0.241&$(0.229, 0.254)$&0.232&$(0.220, 0.245)$\\[0.2cm]
 &100$n$&0.109&0.109&$(0.103, 0.115)$&0.109&$(0.103, 0.115)$\\[0.1cm]
 \hline
\end{tabular}}}
\end{table}

\begin{figure}[h!p!t!]
\centering
\subfigure[Naegleria Aerobic]{
\includegraphics[width=0.48\linewidth]{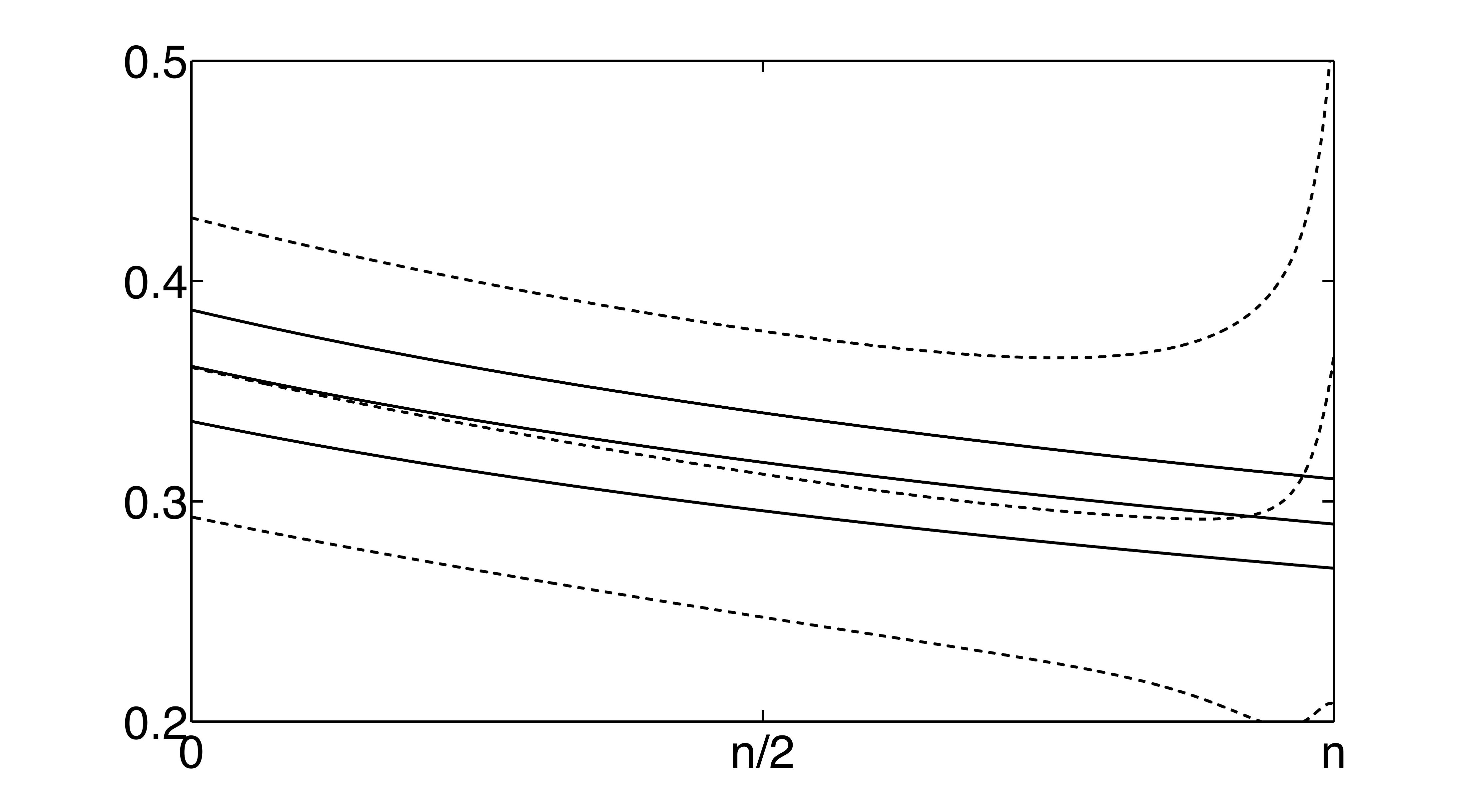}}
\subfigure[Naegleria Anaerobic]{
\includegraphics[width=0.48\linewidth]{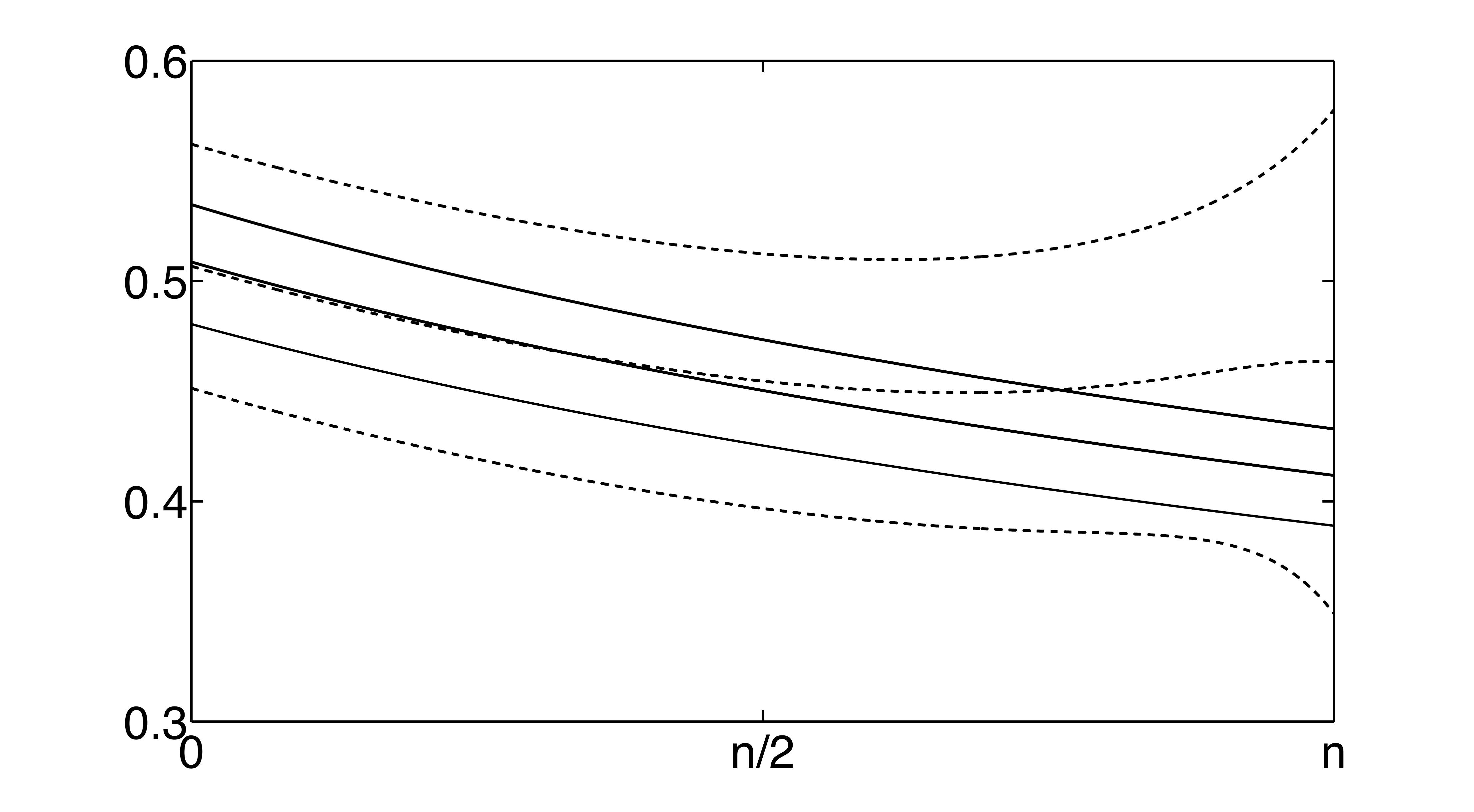}}
\caption{Comparison of Good--Toulmin estimator $\check{\mathcal{D}}_{n,m}(0)$ (inner dashed curves) and Bayesian nonparametric estimator $\hat{\mathcal{D}}_{n,m}(0)$ (inner solid curves) for $m$ ranging in $[0,n]$. The Good--Toulmin estimates are endowed with $95\%$ confidence intervals (outer dashed curves). Bayesian nonparametric estimators are endowed with asymptotic $95\%$ credible intervals (outer solid curves).}
\label{fig:HPD_CI}
\end{figure}

The estimator $\hat{\mathcal{D}}_{n,m}(0)$ is compared with the Good--Toulmin estimator $\check{\mathcal{D}}_{n,m}(0)$. Confidence intervals for  $\check{\mathcal{D}}_{n,m}(0)$, which have been devised in \citet{Mao(04)} via a moment-based approach, and asymptotic credible intervals for $\hat{\mathcal{D}}_{n,m}(0)$ are also compared. We focus on $m\in[0,n]$: such choice reflects the fact that $\check{\mathcal{D}}_{n,m}(0)$ is known to be a good estimator for small $m$, namely $m\leq n$. See \citet{Mao(04)} for details. Figure \ref{fig:HPD_CI} highlights common features for the estimates obtained for the \emph{Naegleria gruberi} libraries. When $m$ is close to $0$ both the approaches  provide similar estimates for the $(m;0)$-discovery. However, even for small values of $m$, asymptotic credible intervals are narrower than the corresponding moment-based $95\%$ confidence intervals. This difference becomes more substantial when $m$ increases. While the asymptotic credible intervals show a regular behavior around the corresponding point estimates, with intervals that tend to get narrow very slowly, estimates obtained with the Good--Toulmin estimator and corresponding confidence intervals feature a more irregular behaviour. The latter approach can lead to estimates with very different behaviors, as $m$ approaches $n$.

We conclude this section by determining the asymptotic credible intervals for the point estimators $\hat{\mathcal{D}}_{n,m}(l)$ and $\hat{\mathcal{D}}_{n,m}(l_1,\ldots,l_\tau)$, for some choices of $l$, $\tau$ and $\{l_{1},\ldots,l_{\tau}\}$. With regards to the \emph{Naegleria gruberi} libraries, Bayesian nonparametric inference for discovery probabilities have been recently considered in \citet{Fav(09)} and \citet{Fav(12)}, where estimates for discovery probabilities and cumulative discovery probabilities are obtained. However, in \citet{Fav(09)} and \citet{Fav(12)} no measures of uncertainty are provided for these estimates. In Table \ref{table_intervals_1} we summarize estimates of the $(m;l)$-discovery for $l=0,\ldots,4$ and of the $(m;l_{1},\ldots,l_{\tau})$-discovery for $\tau=3,4,5$. These estimates are endowed with asymptotic $95\%$ credible intervals obtained by combining asymptotic results displayed in \eqref{eq:asymp_dist_post} and \eqref{eq:limit_toulmin_cum_pd} with the choice of the scaling factors $r^*(m,l)$ and $r^*(m,l_1,\ldots,l_\tau)$, respectively. Table \ref{table_intervals_1} thus complete the illustrations presented in \citet{Fav(09)} and \citet{Fav(12)}.

\begin{table}[h!p!t!]
\caption{\label{table_intervals_1}\emph{Naegleria} aerobic and \emph{Naegleria} anaerobic libraries. $\hat{\mathcal{D}}_{n,m}(l)$, for $l=0,1,2,3,4$, and $\hat{\mathcal{D}}_{n,m}(0,\ldots,\tau)$, for $\tau=3,4,5$, and corresponding asymptotic $95\%$ credible intervals (c.i.). \medskip}
\centering{
\tabcolsep=0.1cm
\small{
\begin{tabular}{*{8}{c}}
\hline
&  \multirow{2}{*}{Library} & \multicolumn{2}{c}{$m=n$} & \multicolumn{2}{c}{$m=2n$} & \multicolumn{2}{c}{$m=3n$}\\\cline{3-4} \cline{5-6} \cline{7-8}\\[-0.2cm]
& & $\text{estimate}$ & 95\% c.i. & $\text{estimate}$ & 95\% c.i. & $\text{estimate}$ & 95\% c.i.\\[0.1cm]
\hline\\[-0.2cm]
 \multirow{2}{*}{$(m;0)$-discovery} & aerobic & 0.289 & (0.267, 0.312) & 0.253 & (0.234, 0.273) & 0.231 & (0.213, 0.249)\\[0.2cm]
 & anaerobic & 0.409 & (0.387, 0.431) & 0.358 & (0.339, 0.378) & 0.326 & (0.309, 0.344)\medskip\\[0.1cm]
\multirow{2}{*}{$(m;1)$-discovery}& aerobic  & 0.093 & (0.084, 0.101) & 0.083 & (0.076, 0.089) & 0.075 & (0.070, 0.081)\\[0.2cm]
& anaerobic & 0.130 & (0.123, 0.137) & 0.117 & (0.111, 0.124) & 0.108 & (0.102, 0.114)\medskip\\[0.1cm]
\multirow{2}{*}{$(m;2)$-discovery}& aerobic  & 0.061 & (0.057, 0.066) & 0.054 & (0.050, 0.059) & 0.050 & (0.046, 0.054)\\[0.2cm]
& anaerobic & 0.080 & (0.076, 0.085) & 0.075 & (0.071, 0.079) & 0.070 & (0.066, 0.074)\medskip\\[0.1cm]
\multirow{2}{*}{$(m;3)$-discovery} & aerobic & 0.046 & (0.042, 0.049) & 0.041 & (0.038, 0.045) & 0.038 & (0.035, 0.041)\\[0.2cm]
& anaerobic & 0.059 & (0.056, 0.062) & 0.055 & (0.052, 0.058) & 0.052 & (0.050, 0.055)\medskip\\[0.1cm]
\multirow{2}{*}{$(m;4)$-discovery} & aerobic & 0.036 & (0.033, 0.039) & 0.034 & (0.031, 0.036) & 0.031 & (0.029, 0.034)\\[0.2cm]
& anaerobic & 0.045 & (0.042, 0.047) & 0.044 & (0.042, 0.046) & 0.042 & (0.040, 0.044)\medskip\\[0.1cm]
\multirow{2}{*}{$(m;0,1,2,3)$-discovery}& aerobic  & 0.490 & (0.452, 0.528) & 0.432 & (0.399, 0.465) & 0.394 & (0.364, 0.425)\\[0.2cm]
& anaerobic & 0.679 & (0.642, 0.716) & 0.606 & (0.573, 0.640) & 0.556 & (0.526, 0.587)\medskip\\[0.1cm]
\multirow{2}{*}{$(m;0,1,2,3,4)$-discovery}& aerobic  & 0.526 & (0.485, 0.563) & 0.465 & (0.430, 0.501) & 0.425 & (0.393, 0.459)\\[0.2cm]
& anaerobic & 0.724 & (0.685, 0.763) & 0.650 & (0.615, 0.686) & 0.599 & (0.566, 0.631)\medskip\\[0.1cm]
\multirow{2}{*}{$(m;0,1,2,3,4,5)$-discovery} & aerobic & 0.556 & (0.514, 0.599) & 0.494 & (0.456, 0.532) & 0.452 & (0.418, 0.487)\\[0.2cm]
& anaerobic & 0.760 & (0.718, 0.801) & 0.686 & (0.649, 0.723) & 0.634 & (0.599, 0.668)\\[0.1cm]
\hline
\end{tabular}}}
\end{table}


\appendix

\section{Appendix}

This appendix contains: i) the proofs of Theorems 1, Theorem 2 and Proposition 1; ii) the explicit expressions for the alternative scaling factors $r^*(m,l)$ and $r^*(m,l_1,\ldots,l_\tau)$; iii) details on the fast rejection sampling by \citet{Hof(11)}; iv) a sensitivity analysis for the asymptotic credible intervals with respect to the choice of the parameter $(\sigma,\theta)$.


\subsection{Proofs}

The proof of Theorem 1 relies on the large $n$ asymptotic behaviours of $K_{n}$ and $M_{l,n}$. For any $\sigma\in(0,1)$ let $f_{\sigma}$ be the density function of the positive $\sigma$-stable random variable. We introduce a random variable $Z_{\sigma,q}$, for any real $q>-1$, with density function $f_{Z_{\sigma,q}}(z)=\Gamma(q\sigma+1)z^{q-1-1/\sigma}f_{\sigma}(z^{-1/\sigma})/\sigma\Gamma(q+1)$. The random variable $Z^{-1/\sigma}_{\sigma,q}$ is referred to as the polynomially tilted positive $\sigma$-stable random variable. Theorem 3.8 and Lemma 3.11 in \citet{Pit(06)} showed that, as $n\rightarrow+\infty$
\begin{equation}\label{eq:asymp_dist}
\frac{K_{n}}{n^{\sigma}}\stackrel{\text{a.s.}}{\longrightarrow}Z_{\sigma,\theta/\sigma}.
\end{equation}
and
\begin{equation}\label{eq:asymp_dist_freq}
\frac{M_{l,n}}{n^{\sigma}}\stackrel{\text{a.s.}}{\longrightarrow}\frac{\sigma(1-\sigma)_{l-1}}{l!}Z_{\sigma,\theta/\sigma}.
\end{equation}
In other terms, according to the fluctuation limits \eqref{eq:asymp_dist} and \eqref{eq:asymp_dist_freq}, as $n$ tends to infinity the number of species with frequency $l$ in a sample of size $n$ from $P_{\sigma, \theta}$ becomes, almost surely, a proportion $\sigma(1-\sigma)_{l-1}/l!$ of the number of species in a sample of size $n$ from $P_{\sigma, \theta}$. The reader is referred to \citet{Pit(06)} and to \citet{Gne(07)} for additional details and refinements of \eqref{eq:asymp_dist} and \eqref{eq:asymp_dist_freq}.

\medskip

\textsc{Proof of Theorem 1.} 
Let us define $c_{\sigma,l}=\sigma(1-\sigma)_{l-1}/l!$, and observe that for $m=0$ the estimators $\hat{\mathcal{D}}_{n,m}(0)$ and $\hat{\mathcal{D}}_{n,m}(l)$ reduce to $(\theta+\sigma k_{n})/(\theta+n)$ and $(l-\sigma)m_{l,n}/(\theta+n)$, respectively. The proof follows by combining the predictive distribution of $P_{\sigma,\theta}$ with the fluctuation limits \eqref{eq:asymp_dist} and \eqref{eq:asymp_dist_freq}. Specifically, let $(\Omega,\mathcal{F},\P)$ be the probability space in which the sample $\boldsymbol{X}_{n}$ is defined. Then, for any $\omega\in\Omega$, a version of the predictive distribution of $P_{\sigma,\theta}$ corresponds to
\begin{displaymath}
\frac{\theta+\sigma K_{n}(\omega)}{\theta+n}\nu_{0}(\cdot)+\frac{1}{\theta+n}\sum_{i=1}^{K_{n}(\omega)}(N_{i,n}(\omega)-\sigma)\delta_{X_{i}^{\ast}(\omega)}(\cdot).
\end{displaymath} 
According to \eqref{eq:asymp_dist} and \eqref{eq:asymp_dist_freq}, $\lim_{n\rightarrow+\infty}c_{\sigma,l}M_{l,n}/K_{n}=1$ almost surely. See Lemma 3.11 in \citet{Pit(06)} for additional details. Since $(\theta+\sigma K_{n})/(\theta+n)\stackrel{\text{a.s.}}{\simeq}\sigma K_{n}/n$ and $M_{1,n}\stackrel{\text{a.s.}}{\simeq}\sigma K_{n}$, as $n\rightarrow+\infty$,  a version of the Bayesian nonparametric estimator of the $(0,0)$-discovery coincides with
\begin{equation}\label{eq_vers_stim1}
\frac{\theta+\sigma K_{n}(\omega)}{\theta+n}\simeq\frac{\sigma K_{n}(\omega)}{n}\simeq\frac{M_{1,n}(\omega)}{n},
\end{equation}
as $n\rightarrow+\infty$. For any $l\geq1$, since $(l-\sigma)M_{l,n}/(\theta+n)\stackrel{\text{a.s.}}{\simeq}(l-\sigma)M_{l,n}/n$ and  $M_{l,n}\stackrel{\text{a.s.}}{\simeq}c_{\sigma,l} K_{n}$, as $n\rightarrow+\infty$, a version of the Bayesian nonparametric estimator of the $(0,l)$-discovery coincides with
\begin{equation}\label{eq_vers_stim2}
(l-\sigma)\frac{M_{l,n}(\omega)}{\theta+n}\simeq(l-\sigma)\frac{M_{l,n}(\omega)}{n}\simeq c_{\sigma,l}(l-\sigma)\frac{K_{n}(\omega)}{n}\simeq(l+1)\frac{M_{l+1,n}(\omega)}{n},
\end{equation}
as $n\rightarrow+\infty$. Let us define $\{\omega\in\Omega : \lim_{n\rightarrow+\infty}n^{-\sigma}K_{n}(w)=Z_{\sigma,\theta/\sigma}(\omega),\lim_{n\rightarrow+\infty}n^{-\sigma}M_{l,n}(\omega)=c_{\sigma,l}Z_{\sigma,\theta/\sigma}(\omega)\}=\Omega_{0}$. From the fluctuation limits \eqref{eq:asymp_dist} and \eqref{eq:asymp_dist_freq} we have $\P[\Omega_{0}]=1$. Fix $\omega\in\Omega_{0}$ and denote by $k_{n}=K_{n}(\omega)$ and $m_{l,n}=M_{l,n}(\omega)$ the number of species generated and the number of species with frequency $l$ generated by the sample $\boldsymbol{X}_{n}(\omega)$. Accordingly, the large $n$ asymptotic equivalences stated in theorem follows from \eqref{eq_vers_stim1} and \eqref{eq_vers_stim2}, and the proof is completed.\hfill \qed

\medskip

For any $n\geq1$, the proof of Theorem 2 relies on the large $m$ asymptotic behaviours of $K_{m}^{(n)}\,|\,K_{n}$ and $M_{l,m}^{(n)}\,|\,(K_{n},M_{1,n},\ldots,M_{l,n})$. Recall that the posterior distribution of $D_{0,m}^{(n)}$ and $D_{l,m}^{(n)}$ are related to the distribution of $K_{m}^{(n)}\,|\,K_{n}$ and $M_{l,m}^{(n)}\,|\,(K_{n},M_{1,n},\ldots,M_{l,n})$ via the identities
\begin{equation}\label{eq_id_post_1}
D_{0,m}^{(n)}\,|\,\boldsymbol{X}_{n}\stackrel{\text{d}}{=}\frac{\theta+\sigma K_{n}+\sigma K_{m}^{(n)}}{\theta+n+m}\,|\,\boldsymbol{X}_{n}
\end{equation}
and
\begin{align}\label{eq_id_post_2}
&D_{l,m}^{(n)}\,|\,\boldsymbol{X}_{n}\stackrel{\text{d}}{=}(l-\sigma)\frac{M_{l,m}^{(n)}}{\theta+n+m}\,|\,\boldsymbol{X}_{n},
\end{align}
respectively. The distribution of the random variables $K_{m}^{(n)}\,|\,K_{n}$ and $M_{l,m}^{(n)}\,|\,(K_{n},M_{1,n},\ldots,M_{l,n})$ have been obtained in \citet{Fav(09)} and \citet{Fav(13)}, respectively. Specifically, for any $x=0,1,\ldots,m$,
\begin{align*}
&\notag\P[K_{m}^{(n)}=x\,|\,\boldsymbol{X}_{n}]\\
&\notag\quad=\P[K_{m}^{(n)}=x\,|\,K_{n}=k_{n}]\\
&\notag\quad=\frac{(\theta/\sigma+k_{n})_{x}}{(\theta+n)_{m}}\mathscr{C}(m,x;\sigma,-n+\sigma k_{n}),
\end{align*}
and for any $x=0,1,\ldots,n+m$,
\begin{align*}
&\P[M_{l,m}^{(n)}=x\,|\,\boldsymbol{X}_{n}]\\
&\notag\quad=\P[M_{l,m}^{(n)}=x\,|\,K_{n}=k_{n},(M_{1,n},\ldots,M_{l,n})=(m_{1,n},\ldots,m_{l,n})]\\
&\notag\quad=\frac{1}{x!}\sum_{t=0}^{k_{n}}t!\sum_{(c_{1},\ldots,c_{t})\in\mathcal{C}_{k_{n},t}}\prod_{i=1}^{t}(n_{c_{i},n}-\sigma)_{l-n_{c_{i},n}}\\
&\notag\quad\quad\times\sum_{y=0}^{\lfloor\frac{m+\sum_{i=1}^{t}n_{c_{i},n}}{l}-x\rfloor}\frac{(-1)^{y}}{y!}{x+y\choose t}{m\choose l,\ldots,l,l-n_{c_{1},n},\ldots,l-n_{c_{t},n},m-(x+y)l+\sum_{i=1}^{t}n_{c_{i},n}  }\\
&\notag\quad\quad\quad\times(\sigma(1-\sigma)_{l-1})^{x+y-t}\frac{\left(\frac{\theta}{\sigma}+k_{n}\right)_{x+y-t}(\theta+(x+y)\sigma+n-\sum_{i=1}^{t}n_{c_{i},n})_{m-(x+y)l+\sum_{i=1}^{t}n_{c_{i},n}}}{(\theta+n)_{m}},
\end{align*}
where $\mathscr{C}(m,x;\sigma,\gamma)=(x!)^{-1}\sum_{0\leq i\leq x}(-1)^{i}{x\choose i}(-i\sigma-\gamma)_{m}$ denotes the noncentral generalized factorial coefficient introduced in \citet{Cha(05)}, $\mathcal{C}_{k_{n},t}$ denotes the set of the combinations of size $t$ (without any repetitions) of $\{1,\ldots,k_{n}\}$, and $\lfloor x\rfloor$ stands for the integer part of $x$.

For any $j\leq n$ let $Z_{\sigma,\theta,j}^{(n)}\stackrel{\text{d}}{=}B_{j+\theta/\sigma,n/\sigma-j}Z_{\sigma,(\theta+n)/\sigma}$ where $B_{a,b}$ is a Beta random variable with parameter $(a,b)$ and $Z^{-1/\sigma}_{\sigma,q}$ is a polynomially tilted positive $\sigma$-stable random variable, independent of $B_{a,b}$. Then, according to Proposition 2 in \citet{Fav(09)},  as $m\rightarrow+\infty$ one has
\begin{equation}\label{eq:asymp_dist_post_1}
\frac{K_{m}^{(n)}}{m^{\sigma}}\,|\,(K_{n}=k_{n})\stackrel{\text{a.s.}}{\longrightarrow}Z^{(n)}_{\sigma,\theta,k_{n}}.
\end{equation}
The large $m$ asymptotic behaviour of $M_{l,m}^{(n)}\,|\,(K_{n}=k_{n},(M_{1,n},\ldots,M_{l,n})=(m_{1,n},\ldots,m_{l,n}))$ follows by combining \eqref{eq:asymp_dist_post_1} with Corollary 21 in \citet{Gne(07)}. Specifically, as $m\rightarrow+\infty$, one has
\begin{equation}\label{eq:asymp_dist_freq_post}
\frac{M_{l,m}^{(n)}}{m^{\sigma}}\,|\,(K_{n}=k_{n},(M_{1,n},\ldots,M_{l,n})=(m_{1,n},\ldots,m_{l,n}))\stackrel{\text{a.s.}}{\longrightarrow}\frac{\sigma(1-\sigma)_{l-1}}{l!}Z_{\sigma,\theta,k_{n}}^{(n)}. 
\end{equation}
The fluctuation limits \eqref{eq:asymp_dist_post_1} and \eqref{eq:asymp_dist_freq_post} provide posterior counterparts of \eqref{eq:asymp_dist} and \eqref{eq:asymp_dist_freq}, respectively. In particular, as for the fluctuation limits \eqref{eq:asymp_dist} and \eqref{eq:asymp_dist_freq},  as $m$ tends to infinity the conditional number of species with frequency $l$ in the enlarged sample from $P_{\sigma,\theta}$ becomes, almost surely, a proportion $\sigma(1-\sigma)_{l-1}/l!$ of the conditional number of new species in the additional sample from $P_{\sigma,\theta}$.

\medskip

\textsc{Proof of Theorem 2.}
By exploiting the fluctuation limits  \eqref{eq:asymp_dist_post_1} and \eqref{eq:asymp_dist_freq_post}, the proof is along lines similar to the proof of Theorem 1. In particular, let us define $c_{\sigma,l}=\sigma(1-\sigma)_{l-1}/l!$.  From the fluctuation limits  \eqref{eq:asymp_dist_post_1} and \eqref{eq:asymp_dist_freq_post}, it can be easily verified that $M_{1,m}^{(n)}\,|\,(K_{n}=k_{n}, M_{1,n}=m_{1,n})\stackrel{\text{a.s.}}{\simeq}\sigma K_{m}^{(n)}\,|\,(K_{n}=k_{n})$, as $m\rightarrow+\infty$. This equivalence, combined with \eqref{eq_id_post_1} leads to the following
\begin{equation}\label{eq_vers_stim1_posterior}
D_{0,m}^{(n)}\,|\,\boldsymbol{X}_{n}\stackrel{\text{a.s.}}{\simeq}\frac{\sigma K_{m}^{(n)}}{m}\,|\,(K_{n}=k_{n})\stackrel{\text{a.s.}}{\simeq}\frac{M_{1,m}^{(n)}}{m}\,|\,(K_{n}=k_{n},M_{1,n}=m_{1,n})
\end{equation}
as $m\rightarrow+\infty$. For any $l\geq 1$, from \eqref{eq:asymp_dist_post_1} and \eqref{eq:asymp_dist_freq_post} one has $M_{l,m}^{(n)}\,|\,(K_{n}=k_{n},(M_{1,n},\ldots,M_{l,n})=(m_{1,n},\ldots,m_{l,n}))\stackrel{\text{a.s.}}{\simeq}c_{\sigma,l}K_{m}^{(n)}\,|\,(K_{n}=k_{n})$, as $m\rightarrow+\infty$. This equivalence, combined with \eqref{eq_id_post_2} leads to
\begin{align}\label{eq_vers_stim22}
D_{l,m}^{(n)}\,|\,\boldsymbol{X}_{n}&\stackrel{\text{a.s.}}{\simeq}(l-\sigma)\frac{M_{l,m}^{(n)}}{m}\,|\,(K_{n}=k_{n},(M_{1,n},\ldots,M_{l,n})=(m_{1,n},\ldots,m_{l,n}))\\
&\notag\stackrel{\text{a.s.}}{\simeq}c_{\sigma,l}(l-\sigma)\frac{K_{m}^{(n)}}{m}\,|\,(K_{n}=k_{n})\\
&\notag\stackrel{\text{a.s.}}{\simeq}(l+1)\frac{M_{l+1,m}^{(n)}}{m}\,|\,(K_{n}=k_{n},(M_{1,n},\ldots,M_{l+1,n})=(m_{1,n},\ldots,m_{l+1,n}))
\end{align}
as $m\rightarrow+\infty$. Finally, let us recall that $\hat{\mathcal{D}}_{n,m}(l)=\E[D_{l,m}^{(n)}\,|\,\boldsymbol{X}_{n}]$, $\hat{\mathcal{K}}_{n,m}=\E[K_{m}^{(n)}\,|\,K_{n}=k_{n}]$ and $\hat{\mathcal{M}}_{n,m}(l)=\E[M_{l,m}^{(n)}\,|\,K_{n}=k_{n},(M_{1,n},\ldots,M_{l,n})=(m_{1,n},\ldots,m_{l,n})]$, for any $l=1,\ldots,n+m$. Accordingly, the large $m$ asymptotic equivalences stated in the theorem follows by taking the expected value of both sides of the asymptotic equivalences \eqref{eq_vers_stim1_posterior} and \eqref{eq_vers_stim22}, and the proof is completed.\hfill \qed

\medskip

\textsc{Proof of Proposition 1.}
The fluctuation limit arises from the study of the large $m$ asymptotic behaviour of $\E[\prod_{1\leq i\leq\tau}(M_{l_{i}})^{r_{i}}\,|\,\boldsymbol{X}_{n}]$ with $r_{i}\geq0$ for any $i=1,\ldots,\tau$. In particular, by combining the definition of $D_{(l_{1},\ldots,l_{\tau}),m}^{(n)}\,|\,\boldsymbol{X}_{n}$ with the multinomial theorem, we can write the identity 
\begin{align}\label{main_prop}
&\E\left[\left(\frac{D_{(l_{1},\ldots,l_{\tau}),m}^{(n)}}{m^{\sigma-1}}\right)^{r}\,|\,\boldsymbol{X}_{n}\right]=\sum_{(r_{1},\ldots,r_{\tau})\in\mathcal{P}_{r,\tau}}{r\choose r_{1},\ldots,r_{\tau}}\E\left[\prod_{i=1}^{\tau}\left(\frac{D_{l_{i},m}^{(n)}}{m^{\sigma-1}}\right)^{r_{i}}\,|\,\boldsymbol{X}_{n}\right]
\end{align}
where we set $\mathcal{P}_{r,\tau}=\{(r_{1},\ldots,r_{\tau}):r_{i}\geq 0\text{ for }i=1,\ldots,\tau\text{ and }\sum_{1\leq i\leq \tau}r_{i}=r\}$. Recall that, as $m\rightarrow+\infty$, $m^{-\sigma+1}D_{l,m}^{(n)}\stackrel{\text{a.s.}}{\longrightarrow}\sigma(1-\sigma)_{l}Z_{\sigma,\theta,k_{n}}^{(n)}/l!$. Therefore, the righthand side of \eqref{main_prop} can be approximated by 
\begin{displaymath}
\sum_{(r_{1},\ldots,r_{\tau})\in\mathcal{P}_{r,\tau}}{r\choose r_{1},\ldots,r_{\tau}}\prod_{i=1}^{\tau}\frac{(l_{i}+1)^{r_{i}}}{m^{\sigma r_{i}}}\E\left[\prod_{i=1}^{\tau} (M^{(n)}_{l_{i}+1,m})^{r_{i}}\,|\,\boldsymbol{X}_{n}\right],
\end{displaymath}
where an explicit expression for the mixed moment $\E[\prod_{1\leq i\leq \tau} (M^{(n)}_{l_{i}+1,m})^{r_{i}}\,|\,\boldsymbol{X}_{n}]$ is provided by Corollary 5 in \citet{Ces(12)}. The fluctuation limit, then, follows by a direct application of the standard Stirling approximation $\Gamma(x+y)/\Gamma(x)\simeq x^{y}$ as $x\rightarrow+\infty$, and the proof is completed.\hfill\qed


\subsection{Scaling factors $r^*(m,l)$ and $r^*(m,l_1,\ldots,l_\tau)$}
We provide an explicit expression for the scaling factors $r^*(m,l)$ and $r^*(m,l_1,\ldots,l_\tau)$, for any $\tau\geq1$ and $\{l_{1},\ldots,l_{\tau}\}$ such that $l_{i}\in\{0,1,\ldots,n+m\}$ for any $i=1,\ldots,\tau$. Recall that $r^*(m,l)$ is defined as the solution of $\hat{\mathcal{D}}_{n,m}(l)=r^{*}(m,l)\sigma(1-\sigma)_{l}\E[Z_{\sigma,\theta,k_n}^{(n)}]/l!$, which can be easily determined since $\hat{\mathcal{D}}_{n,m}(l)$ and $\E[Z_{\sigma,\theta,k_n}^{(n)}]$ have an explicit expression. In particular, one obtains
\begin{align*}
r^*(m,l)&=\frac{\Gamma(\theta+n+\sigma)\Gamma(\theta+n+m+\sigma-l)\Gamma(1-\sigma)l!}{(\theta+\sigma k_{n})\Gamma(\theta+n+m+1)}\\
&\quad\times\sum_{i=0}^l \binom{m}{l-i}\frac{m_{i,n}}{\Gamma(i-\sigma)\Gamma(\theta+n-i+\sigma)}.
\end{align*}
Similarly, the scaling factor $r^*(m,l_1,\ldots,l_\tau)$ is defined as the solution of the more general equation $\hat{\mathcal{D}}_{n,m}(l_1,\ldots,l_\tau)=r^{*}(m,l_1,\ldots,l_\tau)\E[Z_{\sigma,\theta,k_n}^{(n)}]\sum_{1\leq i\leq \tau}\sigma(1-\sigma)_{l_i}/l_i!$, which can be easily determined since the estimator $\hat{\mathcal{D}}_{n,m}(l_1,\ldots,l_\tau)$ has an explicit expression. In particular, one obtains
\begin{align*}
&r^*(m,l_1,\ldots,l_\tau)\\
&\quad=\frac{\Gamma(\theta+n+\sigma)\Gamma(1-\sigma)}{(\theta+\sigma k_{n})\Gamma(\theta+n+m+1)}\sum_{i=1}^{\tau}\Gamma(l_i-\sigma+1)\Gamma(\theta+n+m+\sigma-l_i)  \\
&\quad\quad\times\frac{\sum_{1\leq t\leq l_{i}}\binom{m}{l_i-t}m_{t,n}/\Gamma(t-\sigma)\Gamma(\theta+n+\sigma-t)}{\sum_{1\leq i\leq \tau}\Gamma(1-\sigma+l_i)/l_i!}.
\end{align*}
It can be easily verified that $r^*(m,l)\simeq m^{\sigma-1}$ and $r^*(m,l_1,\ldots,l_\tau)\simeq m^{\sigma-1}$ as $m\rightarrow+\infty$. This is obtained by the standard Stirling approximation $\Gamma(x+y)/\Gamma(x)\simeq x^{y}$ as $x\rightarrow+\infty$. Alternative scaling factors may be determined by comparing high-oder moments of $D_{l,m}^{(n)}\,|\,\boldsymbol{X}_{n}$ and $D_{(l_{1},\ldots,l_{\tau}),m}^{(n)}\,|\,\boldsymbol{X}_{n}$ with corresponding high-oder moments of $Z_{\sigma,\theta,k_n}^{(n)}$. See \citet{Fav(09)} for details.

\subsection{A fast rejection sampling by \citet{Hof(11)}}
In order to sample from the limiting random variable $Z_{\sigma,\theta,j}^{(n)}$, we resorted to the rejection algorithm by  \citet{Hof(11)} for generating random variates from an exponentially tilted $\sigma$-stable distribution. Hereafter we briefly recall the main ideas of this rejection sampling. Conditionally on $U_{(\theta+n)/\sigma}=u$, let $L_{\sigma,(\theta+n)/\sigma}$ be a random variable distributed according to an exponentially tilted $\sigma$-stable distribution with tilting parameter $u$; that is, $L_{\sigma,(\theta+n)/\sigma}$ is a continuous and nonnegative random variable with density function proportional to $\exp\{-ux\}f_\sigma(x)$, where $f_{\sigma}$ denotes the density function of a positive $\sigma$-stable random variable. In order to sample $L_{\sigma,(\theta+n)/\sigma}$ we resort to the fast rejection algorithm by \citet{Hof(11)}. This is an exact sampling algorithm, built upon a standard rejection algorithm, that exploits a clever factorization of the Laplace transform of the exponentially tilted $\sigma$-stable random variable.

The main idea underlying the fast rejection algorithm consists in observing that, 
for any positive integer $r$, the random variable $L_{\sigma,(\theta+n)/\sigma}$ coincides in distribution with the sum of $r$ independent and identically distributed random variables $L_{\sigma,(\theta+n)/\sigma}^{(1)},\ldots,L_{\sigma,(\theta+n)/\sigma}^{(r)}$. Each $L_{\sigma,(\theta+n)/\sigma}^{(i)}$ has density function proportional to $\exp\{-ux\}f_{\sigma}(x r^{1/\sigma})$, that is an exponential tilting of a suitably rescaled $\sigma$-stable density function. A standard rejection algorithm can be used to sample each $L_{\sigma,(\theta+n)/\sigma}^{(i)}$, with $i=1,\ldots,r$. In particular \citet{Hof(11)} suggests to choose the value of $r$ that minimizes the total expected number of rejections. We can summarize the fast rejection algorithm, with reference to the specific problem of sampling $L_{\sigma,(\theta+n)/\sigma}$, by means of the following steps: i) set $r=\max\{1,\text{round}\left(u^\sigma\right)\}$, where \emph{round} denotes the nearest integer function; ii) for $i=1,\ldots,r$, sample $L_{\sigma,(\theta+n)/\sigma}^{(i)}$ by means of a standard rejection sampling with envelope $g(x)=f_{\sigma}(xr^{1/\sigma})\exp\left\{u^{\sigma}/r\right\}$; iii) $L_{\sigma,(\theta+n)/\sigma}=\sum_{1\leq i\leq r} L_{\sigma,(\theta+n)/\sigma}^{(i)}$.


\subsection{Sensitivity analysis}

We perform a sensitivity analysis for the asymptotic credible intervals of the estimator $\hat{\mathcal{D}}_{n,m}(0)$, with respect to the choice of the parameter $(\sigma,\theta)$. We consider the \emph{Naegleria gruberi} aerobic and anaerobic libraries, and, for $(\sigma,\theta)\in\{0.2,0.4,0.6,0.8\}\times\{0.1,1,10,100,1000\}$ and $m\in\{n,10n,100n\}$, we compute asymptotic $95\%$ credible intervals for $\hat{\mathcal{D}}_{n,m}(0)$. Tables~\ref{tab_sens_1} and \ref{tab_sens_2} report these credible intervals together with the asymptotic credible intervals corresponding to specification $(\sigma,\theta)=(\hat\sigma,\hat\theta)$ obtained via the empirical Bayes procedure. A high sensitivity to the values of $(\sigma,\theta)$ is apparent. This is in agreement with the fact that, for both these EST libraries, the posterior distribution of $(\sigma,\theta)$ is rather concentrated around $(\hat\sigma,\hat\theta)$.

\bigskip

\begin{table}[h!p!t!]
\caption{Naegleria Aerobic ($n=959$). A sensitivity analysis for the asymptotic $95\%$ credible intervals for $\hat{\mathcal{D}}_{n,m}(0)$ with respect to the choice of $(\sigma,\theta)$.}
\centering{
\tabcolsep=0.1cm
\small{
\begin{tabular}{*{8}{c}}
\hline
& $\theta\, \backslash\, \sigma$ & 0.2 & 0.4 & 0.6 & 0.8 & $\hat\sigma=0.669$ \\
\hline\\[-0.2cm]
\multirow{5}{*}{$m=n$}& 0.1 & (0.052, 0.062) & (0.120, 0.142) & (0.205, 0.243) & (0.317, 0.369)\\
& 1 & (0.053, 0.063) & (0.121, 0.142) & (0.209, 0.242) & (0.317, 0.369) \\
& 10 & (0.057, 0.068) & (0.125, 0.149) & (0.212, 0.251) & (0.324, 0.377) \\
& 100 & (0.103, 0.117) & (0.173, 0.200) & (0.260, 0.301) & (0.371, 0.420) \\
& 1000 & (0.397, 0.416) & (0.462, 0.491) & (0.538, 0.577) & (0.628, 0.669) \\
& $\hat\theta=46.241$ & & & & & (0.267, 0.312) \\[6pt]
\multirow{5}{*}{$m=10n$}& 0.1 & (0.013, 0.016) & (0.043, 0.051) & (0.104, 0.123) & (0.225, 0.262) \\
& 1 & (0.014, 0.016) & (0.044, 0.051) & (0.106, 0.122) & (0.226, 0.263) \\
& 10 & (0.015, 0.017) & (0.045, 0.054) & (0.108, 0.127) & (0.230, 0.268) \\
& 100 & (0.027, 0.031) & (0.064, 0.074) & (0.134, 0.155) & (0.266, 0.301) \\
& 1000 & (0.132, 0.139) & (0.202, 0.215) & (0.311, 0.333) & (0.477, 0.508) \\
& $\hat\theta= 46.241$ & & & & & (0.153, 0.178)\\[6pt]
\multirow{5}{*}{$m=100n$}& 0.1 & (0.002, 0.003) & (0.011, 0.014) & (0.043, 0.051) & (0.144, 0.168)\\
& 1 & (0.002, 0.003) & (0.012, 0.014) & (0.044, 0.050) & (0.145, 0.169)\\
& 10 & (0.002, 0.003) & (0.012, 0.014) & (0.044, 0.052) & (0.148, 0.172)\\
& 100 & (0.004, 0.005) & (0.017, 0.020) & (0.055, 0.064) & (0.171, 0.194)\\
& 1000 & (0.024, 0.025) & (0.056, 0.060) & (0.132, 0.142) & (0.311, 0.331)\\
& $\hat\theta= 46.241$ & & & & & (0.074, 0.086)\\[6pt]
\hline
\end{tabular}}}
\label{tab_sens_1}
\end{table}

\begin{table}[h!p!t!]
\caption{Naegleria Anaerobic ($n=969$). A sensitivity analysis for the asymptotic $95\%$ credible intervals for $\hat{\mathcal{D}}_{n,m}(0)$ with respect to the choice of $(\sigma,\theta)$.}
\centering{
\tabcolsep=0.1cm
\small{
\begin{tabular}{*{8}{c}}
\hline
& $\theta\, \backslash\, \sigma$ & 0.2 & 0.4 & 0.6 & 0.8 & $\hat\sigma=0.656$ \\
\hline\\[-0.2cm]
\multirow{5}{*}{$m=n$}& 0.1 & (0.070, 0.081) & (0.160, 0.183) & (0.275, 0.316) & (0.428, 0.478)\\
& 1 & (0.070, 0.081) & (0.160, 0.185) & (0.276, 0.316) & (0.429, 0.481)\\
& 10 & (0.074, 0.086) & (0.165, 0.191) & (0.282, 0.320) & (0.433, 0.485)\\
& 100 & (0.119, 0.133) & (0.209, 0.237) & (0.326, 0.367) & (0.471, 0.524) \\
& 1000 & (0.405, 0.426) & (0.485, 0.517) & (0.577, 0.615) & (0.684, 0.726) \\
& $\hat\theta=155.408$ & & & & & (0.387, 0.430) \\[6pt]
\multirow{5}{*}{$m=10n$}& 0.1 & (0.018, 0.021) & (0.057, 0.066) & (0.139, 0.160) & (0.304, 0.340)\\
& 1 & (0.018, 0.021) & (0.058, 0.066)  & (0.140, 0.160) & (0.305, 0.342)\\
& 10 & (0.019, 0.022) & (0.059, 0.069)  & (0.143, 0.162) & (0.308, 0.345)\\
& 100 & (0.032, 0.035) & (0.077, 0.088)  & (0.168, 0.189) & (0.337, 0.376)\\
& 1000 & (0.134, 0.141) & (0.212, 0.226)  & (0.333, 0.354) & (0.519, 0.551)\\
& $\hat\theta=155.408$ & & & & & (0.220, 0.245)\\[6pt]
\multirow{5}{*}{$m=100n$}& 0.1 & (0.003, 0.004) & (0.015, 0.017) & (0.057, 0.066) & (0.195, 0.218)\\
& 1 & (0.003, 0.004) & (0.015, 0.018) & (0.058, 0.066) & (0.196, 0.219)\\
& 10 & (0.003, 0.004) & (0.016, 0.018) & (0.059, 0.067) & (0.198, 0.221)\\
& 100 & (0.005, 0.006) & (0.021, 0.023) & (0.069, 0.078) & (0.217, 0.242)\\
& 1000 & (0.024, 0.026) & (0.059, 0.063) & (0.142, 0.151) & (0.339, 0.359)\\
& $\hat\theta=155.408$ & & & & & (0.103, 0.115)\\[6pt]
\hline
\end{tabular}}}
\label{tab_sens_2}
\end{table}


\section*{Acknowledgments}
The authors are grateful to an Associate Editor and an anonymous referee for their constructive comments and suggestions. Stefano Favaro is supported by the European Research Council through StG N-BNP 306406. Yee Whye Teh is supported by the European Research Council through the European Unions Seventh Framework Programme (FP7/2007-2013) ERC grant agreement 617411.


\end{document}